\documentclass[useAMS,usenatbib]{mn2e}

\DeclareSymbolFont{cmletters}{OML}{cmm}{m}{it}
\DeclareMathSymbol{v}{\mathalpha}{cmletters}{"76}

\usepackage{tabularx,ragged2e,booktabs,caption}
\usepackage{amsmath}
\usepackage{amssymb}
\usepackage{epsfig}
\usepackage{graphicx}
\usepackage{ifthen}
\usepackage{latexsym}
\usepackage{rotating}
\usepackage{subfigure}
\usepackage{times,epsf}
\usepackage{txfonts}
\usepackage{varioref}
\usepackage{verbatim}
\usepackage{array}
\usepackage{url}
\usepackage{color}
\usepackage[dvipsnames]{xcolor}
\usepackage[T1]{fontenc}

\newcolumntype{L}[1]{>{\raggedright\arraybackslash}p{#1}}
\newcolumntype{C}[1]{>{\centering\arraybackslash}p{#1}}
\newcolumntype{R}[1]{>{\raggedleft\arraybackslash}p{#1}}

\newcommand{\be}{\begin{equation}}
\newcommand{\ee}{\end{equation}}
\newcommand{\bea}{\begin{eqnarray}}
\newcommand{\eea}{\end{eqnarray}}

\newcommand\apj{Astrophysical Journal}
\newcommand\apjl{Astrophysical Journal Letters}

\newcommand\apjs{Astrophysical Journal Suppl. Ser.}

\newcommand\aap{Astronomy \& Astrophysics}

\newcommand\mnras{Monthly Notices of the Royal Astronomical Society}

\newcommand\pasj{Publications of the Astronomical Society of Japan}

\newcommand\jqsrt{Journal of Quantitative Spectroscopy and Radiative Transfer}

\newcommand{\avg}[1]{\langle #1\rangle}

\newcommand{\koral}{\texttt{KORAL}\,}
\newcommand{\Medd}{\dot M_{\rm Edd}}
\newcommand{\Ledd}{L_{\rm Edd}}
\newcommand{\medd}{\dot M_{\rm Edd}}
\newcommand{\Msun}{M_\odot}

\title[3D simulations of super-critical disks]{Three-dimensional
  simulations of super-critical black hole accretion disks ---
  luminosities, photon trapping and variability.}
\author[A. S\k{a}dowski, R. Narayan]
       {Aleksander S\k{a}dowski$^1$\footnotemark[1]	        
        and Ramesh Narayan$^{1}$\thanks{E-mail: asadowsk@mit.edu (AS); 
rnarayan@cfa.harvard.edu (RN);} \\
        $^1$ MIT Kavli Institute for Astrophysics and Space Research
77 Massachusetts Ave, Cambridge, MA 02139, USA\\
$^2$ Harvard-Smithsonian Center for Astrophysics, 60 Garden St., Cambridge, MA 02134, USA}

\begin{document}

\maketitle

\label{firstpage}

\begin{abstract}

We present a set of four three-dimensional, general relativistic,
radiation MHD simulations of black hole accretion at super-critical
mass accretion rates, $\dot{M} > \dot{M}_{\rm Edd}$. We use these
simulations to study how disk properties are modified when we vary the
black hole mass, the black hole spin, or the mass accretion rate.  In
the case of a non-rotating black hole, we find that the total
efficiency is of order $3\%\dot M c^2$, approximately a factor of two
less than the efficiency of a standard thin accretion disk. The radiation
flux in the funnel along the axis is highly super-Eddington, but only a small fraction
of the energy released by accretion escapes in this region.  The bulk
of the $3\%\dot M c^2$ of energy emerges farther out in the disk,
either in the form of photospheric emission or as a wind. In the case
of a black hole with a spin parameter of 0.7, we find a larger
efficiency of about $8\%\dot M c^2$. By comparing the relative
importance of advective and diffusive radiation transport, we show
that photon trapping is effective near the equatorial plane. However,
near the disk surface, vertical transport of radiation by diffusion
dominates.  We compare the properties of our fiducial
three-dimensional run with those of an equivalent two-dimensional
axisymmetric model with a mean-field dynamo. The latter simulation
runs nearly 100 times faster than the three-dimensional simulation,
and gives very similar results for time-averaged properties of the
accretion flow. 

\end{abstract}


\section{Introduction}
\label{s.introduction}

Black hole (BH) accretion disks are common in the Universe.  It
appears that virtually every galaxy harbours a
supermassive black hole (SMBH) in its nucleus and it is likely that
every one of these SMBHs has some kind of an accretion
flow. Moreover, just in our own Galaxy, there are probably thousands
of stellar-mass BHs in binaries accreting gas from their companions,
of which a few dozen have been detected in X-rays and are widely
studied. 

Because of the compactness of BHs, accreting gas can liberate
significant amounts of gravitational energy and make the accreting
systems extraordinary luminous. Moreover, magnetic fields near the BH
encourage the extraction of rotational energy of spinning BHs, leading
to formation of powerful relativistic jets.

Early theoretical work on accretion disks was limited to
one-dimensional analytical models. Later, 1+1 and two-dimensional
models with $\alpha$-viscosity were developed. Because accretion flows
are by nature turbulent, such simplified models were not adequate in
most cases. Moreover, they often made strong assumptions, e.g., no
mass loss, constant $\alpha$, zero-torque at the innermost stable
circular orbit (ISCO), which may not be satisfied in real
systems. This motivated the development of techniques for numerically
modeling multi-dimensional turbulent accretion flows.

The key breakthroughs were the identification of the magnetorotational
instability \citep[MRI, ][]{balbushawley-91} and the development of
magnetohydrodynamical (MHD) numerical codes, both Newtonian and
relativistic. Initial efforts were focused on simulating optically
thin disks, as are likely to be present at the lowest accretion
rates. Such systems are relatively simple to simulate because the
radiation is weak and is, moreover, decoupled from the gas. Only in
recent years have more advanced radiation-MHD (RMHD) codes been
developed which can be used to study radiatively luminous systems. The
pioneering initial work was based on Newtonian codes with crude
(flux-limited diffusion) radiative transport
\citep{ohsuga09,ohsuga11}. This was later followed by codes using more
advanced radiative transport schemes, either still in the Newtonian or
special-relativistic approximation
\citep{jiang+12,jiang+14a,ohsugatakahashi-15}, or in general
relativity \citep{sadowski+koral,mckinney+harmrad,fragile+14}.

So far global simulations of optically thick disks that evolve the
radiation field in parallel to gas have been performed
only for super-critical (exceeding the Eddington value, see equation
\ref{e.medd}) accretion rates \footnote{A number of groups
  \citep[e.g.,][]{shafee+08a,schnittman+13,avara+15}
  have performed simulations of thin disks in pure
  hydrodynamical setup implementing arbitrary cooling function.}.
Such disks are geometrically thick and do not require excessive
resolution near the equatorial plane which, so far, makes
self-consistent simulations
of thinner disks too expensive.

In this paper we continue the numerical study of super-critical BH
accretion flows by performing a set of four three-dimensional
simulations, the parameters of which probe different accretion rates,
BH spins, and BH masses. Such a comparative study using a single code
(and set of assumptions) has not yet been performed. In addition, we
compare the properties of our fiducial three-dimensional model with an
equivalent axisymmetrical two-dimensional run which is simulated using
the mean-field dynamo model described in \cite{sadowski+dynamo}.

We begin with a description of the numerical methods
(Section~\ref{s.method}) and details of the five simulations
(Section~\ref{s.sims}). We then discuss the results, focusing on the
luminosities of the simulated disks (Section~\ref{s.luminosities}),
the efficiency of photon trapping (Section~\ref{s.trapping}), and the
variability of the emitted radiation
(Section~\ref{s.variability}). Finally, we assess the strengths and
weaknesses of 2D simulations (Section~\ref{s.axisymmetric}), and
list the conclusions in the Summary (Section~\ref{s.discussion}).

\section{Numerical methods}
\label{s.method}

The simulations described in this paper were performed with the general relativistic radiation magnetohydrodynamical (GRRMHD) code
\texttt{KORAL} \citep{sadowski+koral} which solves the conservation
equations in 
a fixed, arbitrary spacetime using finite-difference methods. The
equations we solve are,
\bea\label{eq.rhocons}
\hspace{1in}(\rho u^\mu)_{;\mu}&=&0,\\\label{eq.tmunucons}
\hspace{1in}(T^\mu_\nu)_{;\mu}&=&G_\nu,\\\label{eq.rmunucons}
\hspace{1in}(R^\mu_\nu)_{;\mu}&=&-G_\nu,
\eea
where $\rho$ is the gas
density in the comoving fluid frame, $u^\mu$ are the components of the gas four-velocity
as measured in the ``lab frame'', $T^\mu_\nu$ is the
MHD stress-energy tensor in this frame,
\be\label{eq.tmunu}
T^\mu_\nu = (\rho+u_{\rm g}+p_{\rm g}+b^2)u^\mu u_\nu + (p_{\rm g}+\frac12b^2)\delta^\mu_\nu-b^\mu b_\nu,
\ee 
$R^\mu_\nu$ is the stress-energy tensor of radiation, and $G_\nu$ is the radiative
four-force describing the interaction between gas and radiation \citep[see][for a more detailed description]{sadowski+koral2}. Here, $u_{\rm g}$ and $p_{\rm g}=(\Gamma-1)u_{\rm g}$ represent the internal energy and pressure of the 
gas in the comoving frame and $b^\mu$ is the magnetic field 4-vector \citep{gammie03}.
The magnetic pressure is $p_{\rm mag}=b^2/2$ in geometrical units. 

The magnetic field is evolved via the induction equation,
\be
\label{eq.Maxi}
\partial_t(\sqrt{-g}B^i)=-\partial_j\left(\sqrt{-g}(b^ju^i-b^iu^j)\right),
\ee
where $B^i$ is the magnetic field three-vector \citep{komissarov-99},
and $\sqrt{-g}$ is the metric determinant.
The divergence-free criterion is enforced using the flux-constrained 
scheme of \cite{toth-00}. 

The radiation field is evolved
through its energy density and flux, and the radiation stress-energy
tensor is closed by means of the M1 closure scheme
\citep{levermore84,sadowski+koral}. The energy exchange between gas
and radiation is by free-free emission/absorption as well as Compton scattering.
The latter is treated in the ``blackbody''
Comptonization approximation as described in \citet{sadowski+comptonization}.

Four of the five simulations described here were performed in three
dimensions (3D), while the fifth was carried out in 2D,
assuming axisymmetry and using the mean-field dynamo model described
in \cite{sadowski+dynamo} with model parameters identical to those
used there.

We use modified Kerr-Shild coordinates with the inner edge of the
domain inside the BH horizon. The simulations were run with a
moderately high resolution of 252 grid cells spaced logarithmically in
radius, 234 grid cells in the polar angle, concentrated towards the
equatorial plane, and 128 cells in azimuth.

All details of the numerical method are given in \cite{sadowski+koral2}.

\section{Simulations}
\label{s.sims}

\subsection{Units}\label{s.units}

We adopt the following definition
for the Eddington mass accretion rate,
\be
\label{e.medd}
\Medd = \frac{L_{\rm Edd}}{\eta c^2},
\ee
where $L_{\rm Edd}=1.25 \times 10^{38}  M/M_{\odot}\,\rm ergs/s$ is the 
Eddington luminosity, $\eta$ is the radiative efficiency of a thin
disk around a black hole with a given spin $a_* \equiv a/M$,
\be
\label{e.eta}
\eta= 1-\sqrt{1-\frac{2}{3\,R_{\rm ISCO}}}, \ee and $r_{\rm ISCO}$
is the radius of the Innermost Stable Circular Orbit (ISCO). According
to this definition, a standard thin, radiatively efficient disk \citep{ss73,novikovthorne73,frankkingraine85} accreting at
$\Medd$ would have a luminosity of $L_{\rm Edd}$. For zero BH spin,
$\Medd = 2.48 \times 10^{18}M/M_{\odot}  \,\rm g/s$.

Hereafter, we use the
gravitational radius $r_{\rm g}=GM/c^2$ as the unit of length, and $r_g/c$
as the unit of time.

\subsection{Initial setup}

Each of the five simulations was initialized as in
\cite{sadowski+koral2}, viz., by starting with the hydrodynamical
equilibrium torus of \cite{penna-limotorus} with the angular momentum
parameters listed in Table~\ref{t.models}, and then redistributing
the pressure between gas and radiation such that local thermal
equilibrium is established with the total pressure unchanged. The
resulting, radiation-pressure supported torus is close to equilibrium.
The initial density was set through the torus entropy parameter and
was adjusted to provide an optically thick torus that would give
super-critical accretion once the simulation has reached steady state.

The initial torii were threaded by weak poloidal magnetic field.  As
each simulation proceeded, the field grew in strength and led to the
onset of the magnetorotational instability, which triggered and
maintained MHD turbulence in the disk. For most models we started with
multiple loops with alternating polarity. For one model
(\texttt{r011}) we used a single loop. Both field configurations were initialized as in
\cite{sadowski+dynamo}.

We performed two simulations with BH mass $10\Msun$ and zero BH
spin. One of these (run \texttt{r001}) resulted in a mean accretion
rate of $\sim10\medd$, while the other (\texttt{r003}), which was
initialized with a more optically thick torus, accreted at $\sim
175\medd$. These two models allowed us to study the role of the mass
accretion rate on disk properties.

The third model (\texttt{r011}) was initialized with a torus similar
to model \texttt{r001}, but we assumed a rotating BH with spin
parameter $a_*=0.7$. This model was the only one of the five that was
initialized with a single poloidal loop of magnetic field. The hope
was that the single loop would lead to a strong magnetic field at the
horizon and would give a magnetically arrested disk
\citep{igu+03,narayan+mad,tch+11,mtb12}.  However, although this
simulation was run up to a time of nearly 15,000, this duration was
still insufficient to reach the MAD limit. Therefore, the magnetic
field at the BH still did not reach the saturated level appropriate to
the MAD state. Nevertheless, by comparing this model with
\texttt{r001}, we were able to study the effect of BH spin\footnote{In
  a recent paper, \cite{mckinney+madrad} present true MAD simulations
  in the radiative super-critical regime. Those represent a different
  class of accretion flows than the ones considered here.}.

In the fourth model (\texttt{r020}), we increased the BH mass to
$1000\Msun$ BH, but kept the mass accretion rate at $\sim
10\medd$. This enabled us to investigate the role of BH mass.

All of the above models were run in 3D. The fifth model (\texttt{d300}) was a 2D
axisymmetric simulation which used the mean-field dynamo model of
\cite{sadowski+dynamo}. This model was initialized with exactly the same
torus configuration as in model \texttt{r001}. The only difference was
that it was evolved in 2D instead of 3D. The purpose of this model
was to assess how well the 2D dynamo model captures the properties of
the 3D model.

\begin{table*}
\centering
\caption{Model parameters}
\label{t.models}

\begin{tabular}{lccccc}
\hline
\hline
&  \texttt{r001} & \texttt{r003} & \texttt{r011} & \texttt{r020} &
\texttt{d300} (2D)\\
\hline
$M_{\rm BH}$ & $10 M_\odot$& $10 M_\odot$& $10 M_\odot$& $1000 M_\odot$ & $10 M_\odot$ \\
$\dot M/\Medd$  &   10.0  & 175.8 & 9.7 & 10.1 & 8.9 \\
$a_*$ &   0.0 &   0.0 &   0.7 &   0.0 &   0.0  \\
$\rho_{0,\rm init}$ &  $4.27\times 10^{-3}$ &  $6.61\times 10^{-2}$ &
$4.68\times 10^{-3}$ &  $3.99\times 10^{-5}$ & $4.27\times 10^{-3}$ \\
$\beta_{\rm max}$  &  10.0 &  10.0 &  10.0 &  10.0 &  10.0 \\
initial $B$ loops  &  multi. &  multi. &  poloidal &  multi. &  multi. \\
$N_R$ x $N_\theta$ x $N_\phi$ &   252 x 234 x 128  &252 x 234 x 128  &252 x 234 x 128  &252 x 234 x 128  &   252 x 234 x 1 \\
$r_{\rm min}$ / $r_{\rm max}$ / $r_0$ / $H_0$  
& 1.85 / 1000 / 0 / 0.6 & 1.85 / 1000 / 0 / 0.6 & 1.85 / 1000 / 0 / 0.6 & 1.85 / 1000 / 0 / 0.6 & 1.85 / 1000 / 0 / 0.6\\
$\xi$ / $r_1$ / $r_2$ / $r_{\rm in}$ & 0.705 / 40 / 1000 / 22.5 &0.705 / 40 / 1000 / 22.5 &0.705 / 40 / 1000 / 10 &0.705 / 40 / 1000 / 22.5 &0.705 / 40 / 1000 / 22.5 \\
$t_{\rm max}$ &  20,000 & 19,000 & 16,100 & 19,200 & 190,000\\
\hline
\hline
\multicolumn{5}{l}{$\dot M$ - average accretion rate}\\
\multicolumn{5}{l}{$\rho_{0,\rm init}$ - maximal density of the
  initial torus in $\rm g/cm^3$}\\
\multicolumn{5}{l}{ $\beta_{\rm max}$ - maximal value
of initial total to magnetic pressure ratio}\\
 \multicolumn{5}{l}{$N_R$ x $N_\theta$ x $N_\phi$ - resolution }\\
\multicolumn{5}{l}{$r_{\rm min}$ / $r_{\rm max}$ / $r_0$ / $H_0$ -
  grid parameters defined in \cite{sadowski+dynamo}}\\
\multicolumn{5}{l}{$\xi$ / $r_1$ / $r_2$ / $r_{\rm in}$ -
  parameters of the initial torus as defined in \cite{penna-limotorus}}\\
\multicolumn{5}{l}{$t_{\rm max}$ - duration of simulation in units of $GM/c^3$ }
\end{tabular}
\end{table*}

\subsection{Accretion flow properties}
\label{s.results}

\begin{figure}
\includegraphics[width=.95\columnwidth]{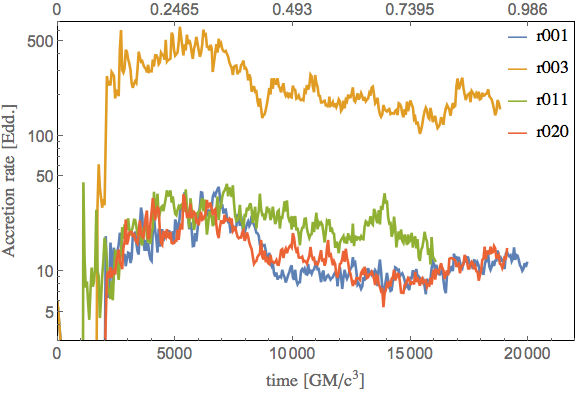}

\caption{Time history of the mass accretion rate at the BH in
  Eddington units (see \S\ref{s.units} for definition) for the four
  three-dimensional models considered in this paper. Model parameters
  are given in Table \ref{t.models}.}
\label{f.mdots}
\end{figure}

Each of the four 3D simulations was run up to a final time $t_{\rm
  max}\approx15,000-20,000$, by which time all had developed
quasi-steady, turbulent accretion via optically and geometrically
thick disks.  The time histories of the mass accretion rate through
the BH horizon are shown in Fig.~\ref{f.mdots}. In all the runs, gas
starts crossing the BH horizon at a substantial rate only after
$t\approx2000$. This is the time needed for the magnetorotational
instability to make the disk turbulent, and for gas from the inner
edge of the initial torus to accrete on the BH. Once
accretion begins, the mass accretion rate increases rapidly. In fact,
$\dot{M}$ overshoots and remains somewhat enhanced until time
$t\approx9000$, and only beyond this time does
the accretion rate settle down to a quasi-steady value.
In the following, we focus on 
disk properties averaged over the latter quasi-steady stage of
accretion, from $t = 9000$ to $t = t_{\rm max}$.

\begin{figure}
\includegraphics[width=.95\columnwidth]{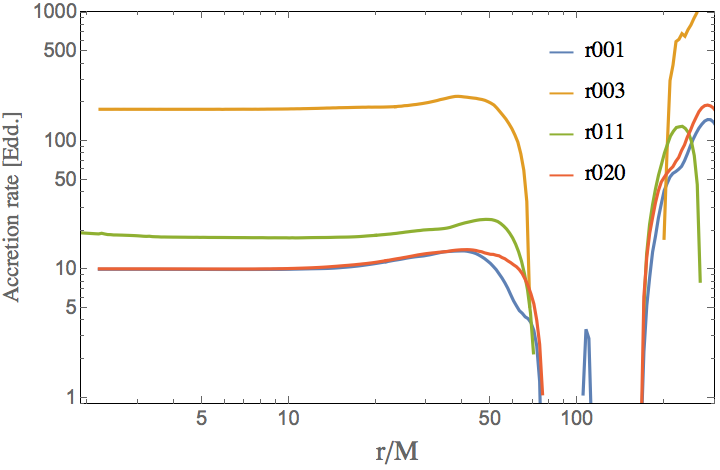}
\caption{Time- and azimuth-averaged radial profiles of the mass
  accretion rates (eq.~\ref{e.mdot}) in the four three-dimensional
  simulations.}
\label{f.mdotsvsr}
\end{figure}

The radial profile of the mass accretion rate is given by 
\be \label{e.mdot} 
\dot M = -\int_{0}^\pi
\int_0^{2\pi}\sqrt{-g}\,\rho u^r{\rm d}\phi {\rm d}\theta.
\ee
Fig.~\ref{f.mdotsvsr} shows the time-averaged profiles of this quantity as
a function of radius $r$ corresponding
to the four 3D runs. The flat sections of the profiles at relatively
small radii denote the region where the flow has reached inflow
equilibrium. Given the somewhat limited duration of the simulations,
and the relatively low radial velocity of the accreting gas in the
disk, inflow equilibrium is reached only up to a radius $r_{\rm
  eq}\sim20-30$. The outflows in the jet and wind regions have larger
velocities, and therefore the outflowing regions are causally
connected with the equatorial disk out to much larger distances from
the BH. In particular, the funnel region, which is filled in most cases with
gas escaping at $v > 0.1c$, achieves equilibrium all the way out to the
domain boundary at $r_{\rm out}=1000$.

\begin{figure*}
\hspace{-.5cm} poloidal plane \hspace{6cm} equatorial plane\\
\includegraphics[width=1.02\columnwidth]{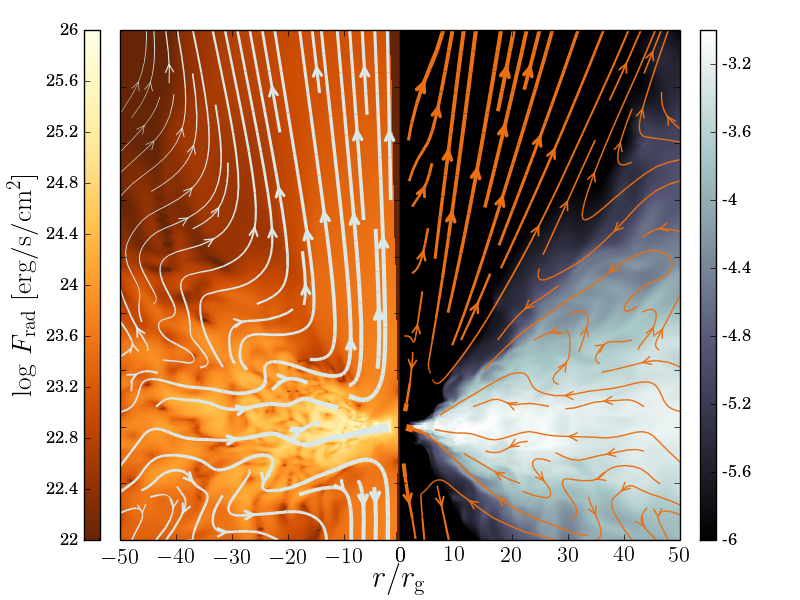}\hspace{-.6cm}
\includegraphics[width=1.02\columnwidth]{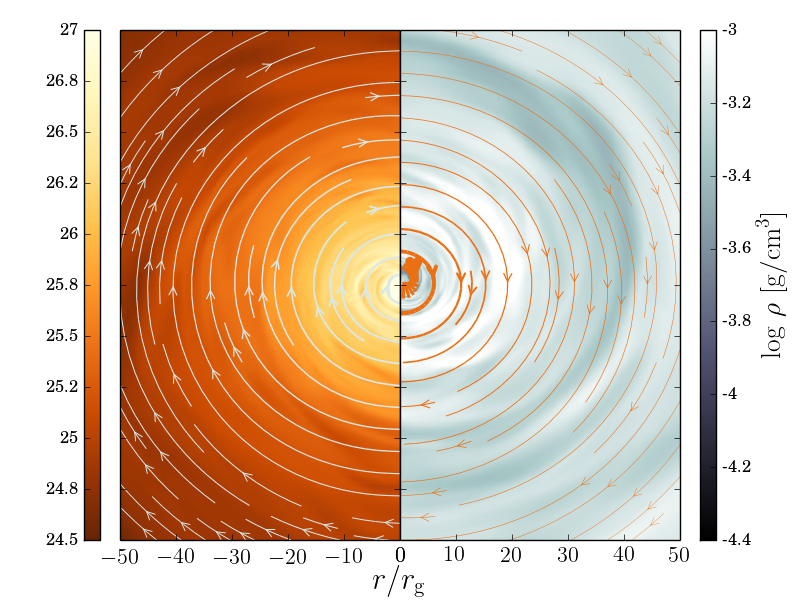}
\caption{Shows the magnitude of the radiative flux (orange colors in
  the left half of each panel) and the gas density (grey colors in the
  right half of each panel) for a snapshot taken near the end of the
  \texttt{r001} simulation. The left and right panels correspond to
  slices through the poloidal and equatorial planes, respectively.
  Streamlines of the radiative flux and gas velocity are azimuth- and
  time-averaged. The thickness of the streamlines increases with the
  magnitude of the respective quantity. }
\label{f.snapplots}
\end{figure*}

Figure~\ref{f.snapplots} shows poloidal and azimuthal slices through
model \texttt{r001} at time $t=16400$. The colors in the left halves
of the panels show the magnitude of the radiation field, while those
in the right halves show the gas density. The arrows in the two halves
show the direction of the radiative flux and gas velocity,
respectively, with the arrow thicknesses indicating the corresponding
magnitudes. For clarity, the vector fields were computed from
time-averaged output.

The gas is
concentrated near the equatorial plane and forms a geometrically thick
(or slim)
disk with density scale-height $H/R\approx 0.25$. Non-axisymmetric
modes are clearly visible in the right panel, showing the value of
running the simulations in 3D. Because of the large density, the gas is
optically thick and advects with it a significant amount of
radiation. This explains why the radiation flux has its largest magnitude
in the bulk of the disk.

The gas in the disk region around the mid-plane is turbulent.  On
average the gas moves there slowly towards the BH. Outside the bulk of
the disk, within $\sim40$ deg from the pole, the gas flows
outward, being driven mostly by the radiation pressure force exerted
by the radial radiative flux\footnote{For a comprehensive study of the
  acceleration mechanism of outflows in hydrodynamical and radiative
  disks see \cite{moller+15}.}.

\begin{figure}
\includegraphics[width=.95\columnwidth]{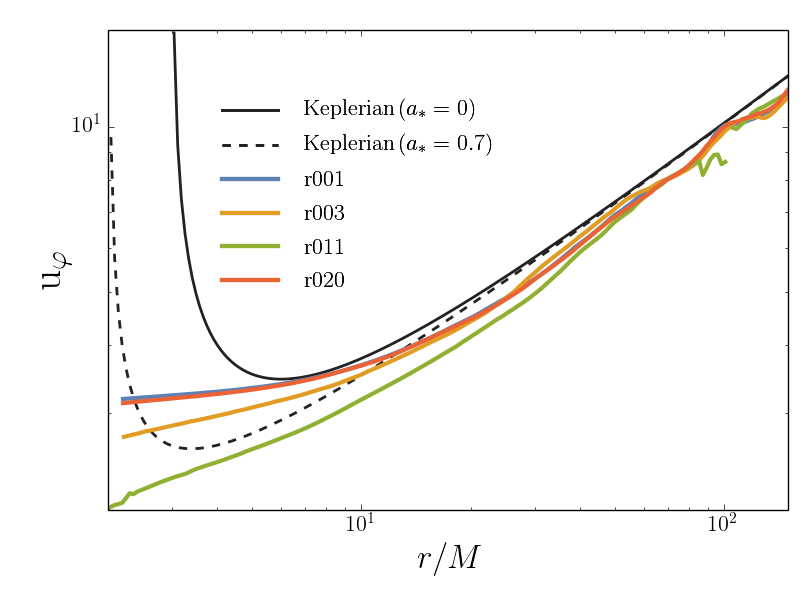}
\caption{Average radial profiles of the density-weighted gas angular
  momentum, $u_\phi$. The black lines show the Keplerian profiles for
  spin $a_*=0.0$ (solid) and $a_*=0.7$ (dashed).}
\label{f.uphivsr} 
\end{figure}

Gas in the disk region orbits around the BH but because the disk is
geometrically thick the rotation is mildly sub-Keplerian, as shown by
Fig.~\ref{f.uphivsr}. Outside the ISCO, the deviation from the
Keplerian profile is no more than 13\% in any of the models. The angular
momentum profile flattens towards the BH horizon, reflecting the fact
that the effective visosity, which transports angular momentum, is
less effective in the plunging region. The profile is flattest for the
models with the lowest accretion rate. Extrapolating to sub-Eddington
accretion rates, we thus expect the viscous torques to become
insignificant for thin disks, in agreement with previous work
\cite{paczynski-thindisks, afshordipaczynski03, shafee+08a, shafee+08b, penna+10}.

The top set of panels in Fig.~\ref{f.5panels} shows the time- and
  azimuth-averaged distribution of density for all the five
  simulations. Streamlines reflect the velocity field of the gas, with the
  thickness of lines denoting the density-weighted average velocity. For all the simulations the accretion flow is
geometrically thick with density peaking at the equatorial plane. The
density values are similar in all the runs except
the simulation with the highest accretion rate,
\texttt{r003}, which has a significantly higher density of gas (and
increased optical depth). In all cases the gas flows on average
towards the BH deep in the disk. The velocity of gas in the funnel is
pointing outward outside the stagnation radius separating the polar region
of inflowing gas (which is the boundary condition imposed by the
presence of BH horizon) and the outflowing gas, driven either by
radiation pressure or pressure gradients. In the case of simulations with
non-rotating BHs and moderate accretion rates (three-dimensional models \texttt{r001}
and \texttt{r020}), the stagnation radius at the axis is located near
$r\sim 10$. For the run with much higher accretion rate
(\texttt{r003}) it shifts significantly outward to $r\sim 50$ due to the much larger
opacity, which prevents radiation from ejecting gas from the
innermost region. On the other hand, the stagnation radius is very close to the BH
horizon for the simulation with a rotating BH. In this model, gas is
accelerated outward by an additional energy source, viz., the spin energy of the BH. The
transition between the region of inflow (inside the disk), and outflow
(in the funnel) is quite rapid --- when gas particles are blown out of
the disk and enter the funnel they quickly gain large outward radial
velocity and join the outflow.

The bottom set of panels in Fig.~\ref{f.5panels} shows the averaged radiative flux for the
five simulations. The colors denote the magnitude 
and arrows show the direction of the radiative flux.
The radiation emitted by hot gas in the disk is mostly advected with the optically
thick gas, but in regions close to the disk surface it also diffuses down the local density gradient \citep{jiang+14b}.
As a result, some fraction of the radiation is advected into
the BH and the rest naturally fills up the funnel region. There, radiation pressure
accelerates gas outward along the axis of the funnel, resulting in a
jet that travels at a modest fraction (0.2--0.5) of the speed of
light \citep{sadowski+radjets}. The properties of the radiation field
depend on the accretion rate and BH spin. For model \texttt{r003}, which has the
largest accretion rate, a significantly higher fraction of all the
radiative flux in the inner region is advected into the BH, even in the
inner part of the funnel region. In simulation \texttt{r011}, the
rotational energy of the BH is extracted (although not very efficiently
due to the sub-MAD level of magnetic flux accumulated at the horizon)
through the Blandford-Znajek process. This extra energy is converted
into radiation and therefore this model has the  highest radiative flux magnitude in the
funnel region among all the runs.

The qualitative properties of the 3D simulations described here agree 
well with accretion flows simulated
in the recent years in axisymmetry e.g., by \cite{sadowski+koral2},
and in three-dimensions by \cite{mckinney+harmrad}. We comment on the differences
between our models and the simulation presented by \cite{jiang+14b} 
in Section~\ref{s.comparison}. Below we discuss
in detail three aspects of our models -- luminosities,
efficiency of photon trapping, and variability.

\section{Luminosities}
\label{s.luminosities}

Accretion takes place if there is an efficient mechanism for transporting
angular momentum outward. In BH accretion disks, we believe that this
transport results from turbulence sustained by
magnetorotational-unstable magnetic field. The angular momentum flux
is accompanied by fluxes of energy in various forms. In the standard
picture of a thin disk, the accreting gas brings in kinetic
(orbital and turbulent), thermal, and binding energy. The latter has a
negative sign, so effectively the flux of binding energy transports
energy out (and deposits it at infinity). The exchange of angular
momentum would not be possible without a shear stress (``viscosity'') which again causes an
outward flux of energy. For turbulent magnetic disks, this energy flux
comes from the work done by magnetic fields. In addition, there is
radiation flux. In a thin disk, radiation carries energy out to infinity. However,
in geometrically thick super-Eddington accretion flows, the radiation energy flux
can be either outward or inward, depending on how effectively
the radiation is
trapped in the optically thick flow. In addition, these models can also have mechanical
energy flowing out in a wind or jet.

We postpone a comprehensive discussion of the energetics in
multi-dimensional accretion disk to a forthcoming paper. Below, we
limit ourselves to simple angle-integrated luminosities of
the simulated disks.

\subsection{Total luminosities}

The most fundamental definition of the luminosity is,
\be \label{e.Ltot}
L_{\rm total}=-
\int_{0}^\pi \int_0^{2\pi}\left(T^r_t+R^r_t+\rho u^r\right)\sqrt{-g}\,{\rm d}\phi {\rm
  d}\theta,
\ee 
which is the total rate of energy flowing through a sphere at some radius $r$ at
some instant of time. This quantity
accounts for all forms of energy except the rest-mass energy (which has been
subtracted out via the term $\rho u^r$). In particular, $L_{\rm total}$
includes binding (gravitational), radiative, kinetic, thermal and
magnetic energies. In an equilibrium steady state accretion disk, the the time-averaged
luminosity $L_{\rm total}$ is conserved, i.e., it has the same value at all
radii. This is because of energy conservation: any
energy that appears in one form must ultimately come from one of the other forms discussed above, and therefore the sum of all forms of energy remains constant.

\begin{figure}
\includegraphics[width=1.0\columnwidth]{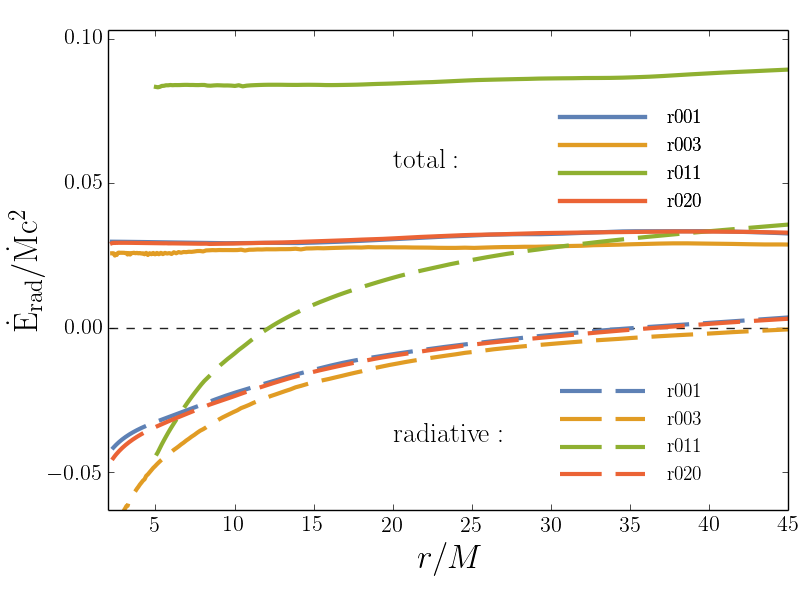}
\caption{Total luminosity (solid lines, eq.~\ref{e.Ltot}) and
  radiative luminosity (dashed lines, eq.~\ref{e.Lradtot}) integrated
  over the whole sphere for the four three-dimensional simulations.}
\label{f.radfluxall}
\end{figure}

The radial profile of total luminosity is plotted with solid lines in
Fig.~\ref{f.radfluxall}. We see that it is indeed constant to good
accuracy for all the runs. All the simulations with non-rotating BHs
have total luminosity close to $3\%\dot Mc^2$. For an accretion rate
of $10\medd$ this corresponds to $\sim 5\Ledd$ (for the adopted
definition of $\medd$ see equation~\ref{e.medd}). The luminosity is
significantly higher $\sim 8\%\dot Mc^2$ for the simulation
(\texttt{r011}) with a rotating BH. There are two reasons for this:
(i) the disk around a spinning BH extends deeper into the potential
well since the ISCO is at a smaller radius (correspondingly, the thin
disk efficiency for $a_*=0.7$ is $\eta_{\rm thin}\approx 10.3\%\dot
Mc^2$), and (ii) the accumulated magnetic flux at the BH horizon
allows for the extraction of BH kinetic energy through the
Blandford-Znajek process. The predicted rate of that process (for the
measured average magnetic flux parameter $\Phi\approx 15$ and
$a_*=0.7$) is $\eta_{\rm jet}\approx 6\%$. The total energy available
is therefore $\eta_{\rm thin}+\eta_{\rm jet}\approx 16\%$. In
actuality, $\sim8\%$ is extracted, which is roughly $50\%$ of the
total available.  A similar 50\% energy extraction is seen also in the
case of non-rotating BHs: a thin disk around a non-spinning BH has an
efficiency of 5.7\% whereas, as mentioned above, the luminosity of our models is only 3\%. Total luminosities 
for all the simulations are given in Eddington
units in the third row of Table~\ref{t.luminosities}.

The total luminosity as defined above may be, in principle, sensitive to
the initial conditions. In the ideal world, one would initiate a
simulation by dumping marginally bound (zero Bernoulli number) gas
from infinity. Because the duration of real simulations is limited, we
have to start our simulatons from a bound torus located close to the BH with its
inner edge at $r_{\rm in}=22.5$. The Bernoulli number, defined as
\be
\label{e.Be}
Be=-\frac{T^t_t+R^t_t+\rho u^t}{\rho u^t}, \ee of the initial gas at
the very inner edge is $Be\approx-0.014$ and approaches zero inversely
proportional to radius. Thus, in principle, the luminosity estimates
given above may be overestimated by up to $\sim 1\%\dot Mc^2$ (it
cannot be more, but it may be less if there is significant radial
mixing of gas in the initial torus). To estimate how strongly the
initial $Be$ affects results we compared the total luminosities
averaged over $t=12,000-13,000$ and $t=18,000-19,000$.  At $t\sim
12,000$, the BH had accreted an amount of gas equivalent to that
contained in the initial torus inside within $r=35$, while at
$t\sim18,000$, the mass accreted was equivalent to the torus mass out
to $r\approx40$. If accretion occurs radius after radius, i.e.,
without any radial mixing, then gas located initially near these radii
should fall on the BH during these periods of time. The initial
Bernoulli numbers at $r=35$ and $r=40$ were $Be=-0.012$ and $-0.010$,
respectively. The meeasured total efficiencies averaged over the
corresponding periods were $\eta_{\rm total}=0.029$ and
$\eta=0.031$. respectively. If the total efficiency of energy
extraction was to reflect the binding energy of the initial gas then
it should \textit{drop} by $0.002$ between the first and second
periods (less bound gas accreted on the BH effecively deposits less
energy at infinity). However, the efficiency \textit{increased} by a
similar amount. We conclude that the gas mixes effectively before
reaching the BH and that the measured total luminosities are not
sensitive to how much the initial torus was bound.

The total luminosity indicates how much energy will be ultimately
deposited at infinity, where binding energy is zero and therefore the
outflowing flux of binding energy must have converted into other forms
of energy. For supermassive black holes
(SMBHs), we may expect that all this energy will ultimately affect the
interstellar medium around the galactic nucleus and will contribute to
AGN feedback.

In analogy with equation (\ref{e.Ltot}), we can straightforwardly
define an equivalent total radiative luminosity, i.e., radiative flux
integrated over the whole sphere, \be \label{e.Lradtot} L_{\rm
  rad,\,total}= -\int_{0}^\pi \int_0^{2\pi}R^r_t\sqrt{-g}\,{\rm d}\phi
{\rm d}\theta, \ee which describes the radiative component of the
total luminosity $L_{\rm total}$. However, this quantity is not as
fundamental as $L_{\rm total}$ since it is no longer conserved. In
particular, radiation can be emitted or absorbed by the gas and can
also gain/lose energy via momentum transfer to the gas. Nevertheless,
for certain limited purposes, $L_{\rm rad,total}$ can be useful. Radial
profiles are shown with dashed lines in
Fig.~\ref{f.radfluxall}. For all the simulations, $L_{\rm rad,total}$ is negative in the
inner region, increases with radius, and ultimately becomes
positive. Negative values correspond to regions where more
photons are dragged (advected) with the flow inward than manage to escape. These
are the regions where the photon-trapping effect (discussed in detail
in Section~\ref{s.trapping}) dominates. The fact that the total
radiative luminosity becomes ultimately positive reflects the fact
that the flow is slower at larger radii and so it is easier for photons
to escape there.

\subsection{Optically thin and outflow regions}

Because of the limited range of inflow/outflow equilibrium in the
disk mid-plane, extending at best only up to
$r_{\rm eq}\sim 25$, it is impossible to determine say how
the outflowing energy is finally distributed between radiation and other forms when
it reaches infinity.
However, the funnel and wind regions
are converged to larger radii because of their higher velocities which allow
them to be causaly connected with the equilibrium innermost region
near the equatorial plane. This fact allows us to measure
luminosities in the funnel to larger distances from the BH.

\begin{figure*}
\includegraphics[width=1.0\textwidth]{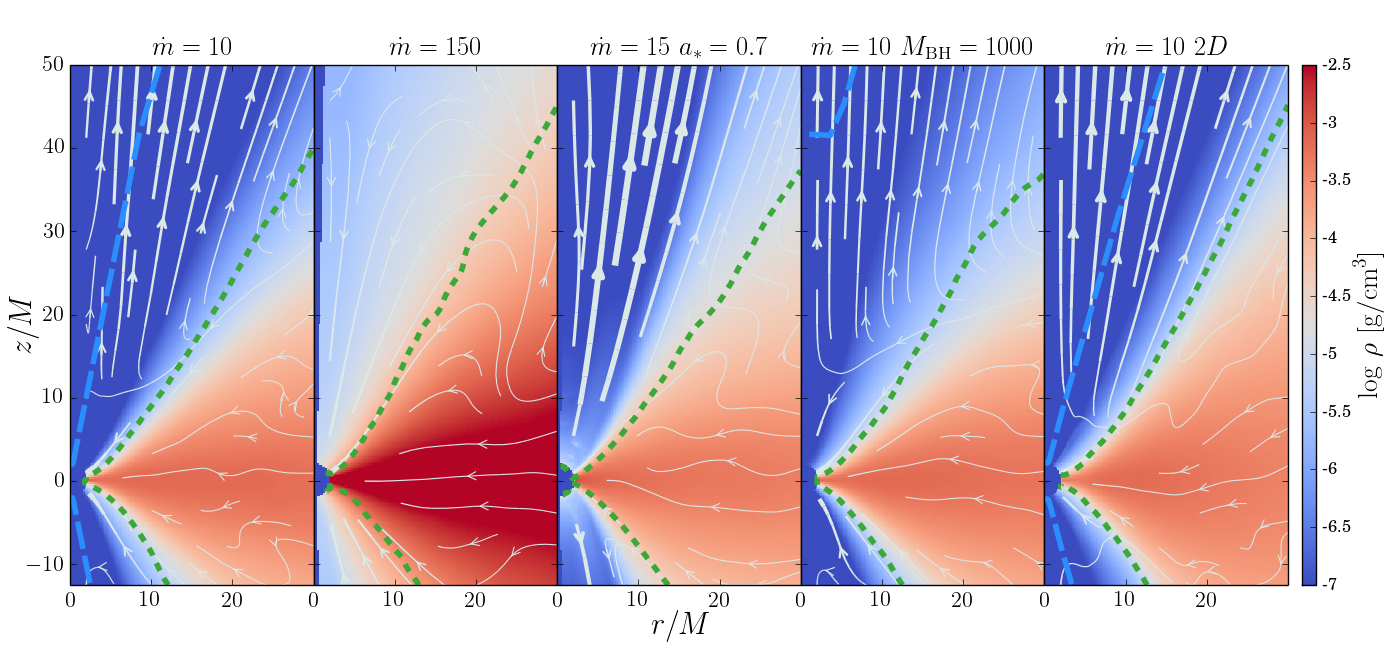}\vspace{-.8cm}
\includegraphics[width=1.0\textwidth]{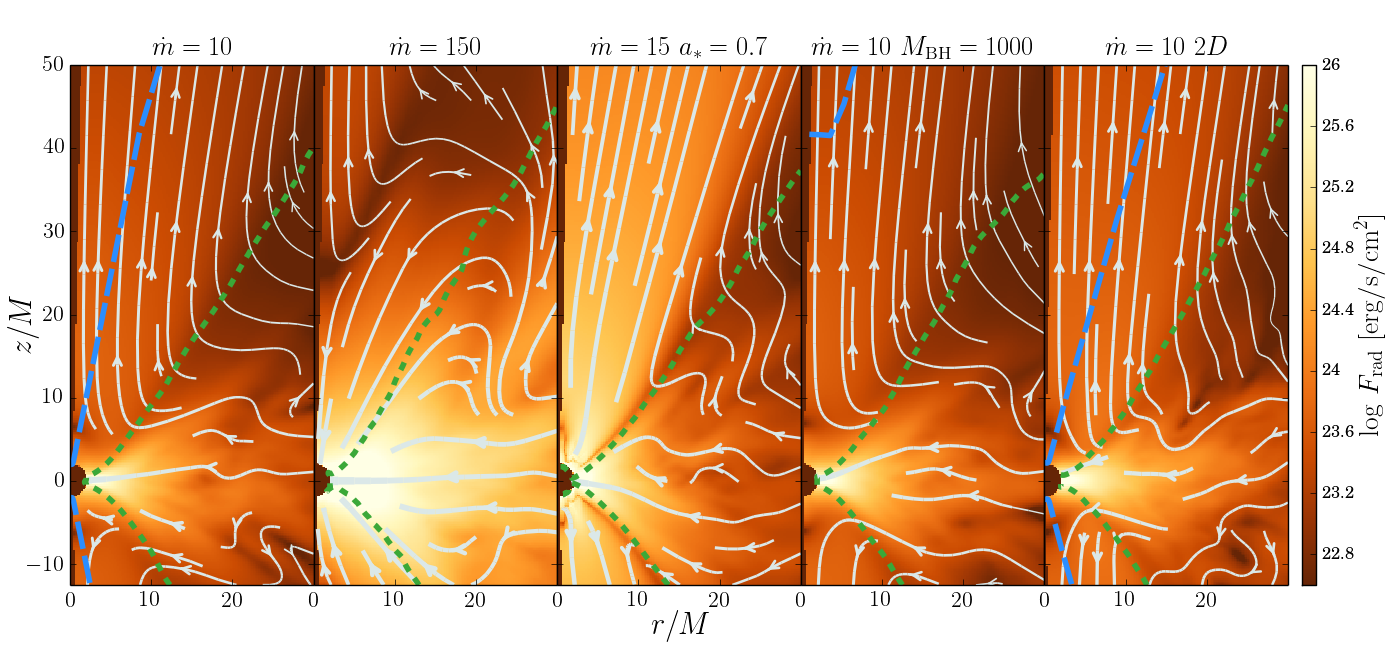}
\caption{\textit{Top row of panels:} Logarithm of average gas density
  (colors) and streamlines of average gas velocity (thickness
  indicates the magnitude of the velocity) for five models:
  \texttt{r001} (leftmost), \texttt{r003}, \texttt{r011},
  \texttt{r020}, and \texttt{d300} (rightmost panel).  Model
  \texttt{r020} corresponds to $M_{\rm BH}=1000 \Msun$), and so its
  density was scaled up by a factor of $100$ to enable direct
  comparison with the other simulations, which had $M_{\rm BH}=10
  \Msun$. The blue dashed contour reflects the location of the
  scattering photosphere as seen from infinity along fixed polar
  angle. The green dashed contour separates the bound ($Be<0$) and
  unbound ($Be>0$) gas. \textit{Bottom row of panels:} Logarithm of
  the magnitude of radiation flux and its streamlines (thickness
  indicates the magnitude of the flux). For model \texttt{r020}, the
  flux was scaled up by $100$.}
\label{f.5panels}
\end{figure*}

We divide the outflow region into two zones: (i) an optically thin
zone which is visible to an observer at infinity, and (ii) an outflow
region where the gas is on average energetic enough to reach
infinity. In all the simulations, zone (i) is a subset of zone (ii).
The border of this zone is defined to be the photosphere, which
satisfies the following condition\footnote{This estimate of the optical depth is for
a light ray propagating radially outward, where we have included the effect of
the motion of the gas but neglected 
gravitational deflection of the ray. We also avoid edge effects, as described
in \cite{sadowski+dynamo}.}, \be
\label{e.photosphere}
\tau(R)=-\int_{R}^{R_{\rm max}}\rho(\kappa_{\rm a}+\kappa_{\rm
  es})(u_t+u_r)\sqrt{g_{rr}}\,{\rm d} r=\frac23, \ee
i.e., the total optical depth integrated along fixed polar angle from
the domain boundary equals $2/3$. The outflow region is defined as the region
where the relativistic Bernoulli parameter (eq.~\ref{e.Be})
is positive. The borders of the two regions are denoted by dashed
blue (optically thin) and green (outflow) lines, respectively, in Fig.~\ref{f.5panels}.
Only for model \texttt{r001}
(and its axisymmetric counterpart \texttt{d300}) does the optically thin
region extend down to the BH. For all the other simulations the density
of gas in the funnel region is large enough to move the lower edge
of the photosphere away from the BH. For model \texttt{r003}, which is
characterized by a significantly larger accretion rate, the photosphere
formally is at the outer edge of the domain. The photosphere is far from the BH
also for model   \texttt{r011} for which the accretion rate (in
Eddington units) is almost twice as high as in the fiducial one. The photosphere is relatively close ($r \approx
40$ at the axis) for model \texttt{r020}, which has a similar accretion
rate as \texttt{r001}, but a higher BH mass. \cite{mckinney+madrad}
has recently showed that if strong magnetic field is present near the axis, one may
get optically thin funnel down to the BH even for highly
super-critical accretion rates. 

Radiation flowing out in the optically thin region is guaranteed to
reach observers at infinity - no significant interaction with gas is
taking place here. The radiative luminosity integrated over this
region may thus be viewed as a lower limit on the total radiative
luminosity. The photons trapped in the optically thick regions can in
principle ultimately escape and could increase the radiative
luminosity. However, this additional luminosity is not expected to
exceed a couple of $\Ledd$, because if locally the radiative flux
exceeds the Eddington flux $\Ledd/c^2$ and the gas is optically thick,
this flux would accelerate the gas and convert radiative energy into
kinetic energy of the outflow\footnote{In principle, since gas in
    the optically thick wind moves radially outward, beaming can allow
    a super-Eddington flux at the wind photosphere, as measured in the
    lab frame. However, the velocities in our simulations are only a
    modest fraction of $c$, so the enhancement factor is unlikely to
    be large.}. 

To obtain the radiative luminosities of the optically thin and outflow zones, we
calculate,
\bea \label{e.Lradthin}
\hspace{.5cm}L_{\rm rad,\,thin}&=&-
\int_{0}^{\theta_{\rm thin}}\int_0^{2\pi}R^r_t\sqrt{-g}\,{\rm d}\phi {\rm
  d}\theta,\\\label{e.Lradout}
L_{\rm rad,\,out}&=&-
\int_{0}^{\theta_{\rm out}}\int_0^{2\pi}R^r_t\sqrt{-g}\,{\rm d}\phi {\rm
  d}\theta,
\eea 
where $\theta_{\rm thin}$ and $\theta_{\rm out}$
denote the limits of the regions of optically thin and outflowing
gas, respectively. The kinetic luminosities are calculated in a similar way
\bea \label{e.Lkinthin}
\hspace{.5cm}L_{\rm kin,\,thin}&=&-
\int_{0}^{\theta_{\rm thin}}\int_0^{2\pi}(u_t+\sqrt{-g_{tt}})\rho u^r\sqrt{-g}\,{\rm d}\phi {\rm
  d}\theta,\\ \label{e.Lkinout}
L_{\rm kin,\,out}&=&-
\int_{0}^{\theta_{\rm out}}\int_0^{2\pi}(u_t+\sqrt{-g_{tt}})\rho u^r\sqrt{-g}\,{\rm d}\phi {\rm
  d}\theta.
\eea 
In the Newtonian limit one has: $-(u_t+\sqrt{-g_{tt}})\rightarrow 1/2v^2$.

Table~\ref{t.luminosities} lists the radiative and kinetic luminosities
in the two regions as measured at radii $r=50$ and $r=250$. For the
fiducial model \texttt{r001} (accreting at $10.0\Medd$) only $0.3\Ledd$ escapes in the optically
thin region at radius $r=50$. This value increases to $0.95\Ledd$ at
radius $r=250$ reflecting the fact that the optically thin region
extends further from the axis and more radiation is able to enter the
optically thin funnel. The radiative luminosity in the whole region of
outflowing ($Be>0$) gas is $1.13\Ledd$ and $1.57\Ledd$ at $r=50$ and
$r=250$, respectively. At the same time kinetic luminosities are much
smaller. Hardly any kinetic energy escapes in the optically thin
region, mostly because of the negligible mass flux of outflowing gas
there. In the whole outflow region, the kinetic luminosity grows from
$0.17\Ledd$ at radius $r=50$ up to $0.35\Ledd$ at $r=250$. These
numbers correspond to $13\%$ and $18\%$ of the total (radiative plus
kinetic) luminosities, respectively. The increase of the fractional
contribution of kinetic energy reflects the fact that radiation
gradually accelerates gas, thereby losing momentum/energy.

This effect is clear especially for simulation $\texttt{r003}$
(accreting at $175\medd$). The fractional contribution of kinetic
luminosity grows from $39\%$ to $58\%$ between $r=50$ and $r=250$. The
transfer of energy from radiation to gas is more effective because of
higher optical depth in this run. For the same reason, there is
no optically thin region in this simulation within the computational
domain. The fact that both luminosities in simulation $\texttt{r003}$
grow significantly between the two radii arises from the fact that
there is strong inflow of gas along the axis within $r\approx 30$
which prevents photons from escaping. Only further out on the axis does the
standard, radiatively driven outflow region form.

The kinetic and radiative luminosities measured in the thin
and outflow regions for runs \texttt{r001} and \texttt{r003} are significantly lower than the total
luminosities measured according to equation~(\ref{e.Ltot}). At radius
$r=250$ and in the outflow region, these luminosities
contribute only $37\%$ and $10\%$ of the total, respectively. Where does the
rest of the luminosity go?  It remains still in the optically thick gas
in the disk interior, where the total energy budget is dominated
by the outflowing flux of magnetic (viscous stresses) and
binding energy. The energy carried out in these channels
(certainly, the binding energy) will be ultimately converted
into radiative or kinetic energy by the time it reaches infinity.
Unfortunately, the limited range of the equilibrium
solution in our simulations prevents us from addressing this question
directly.

The total efficiency in simulation \texttt{r011}, which is the only
one with a rotating BH, is significantly higher than in the other
simulations. So are the radiative and kinetic luminosities integrated
over the outflow region. The extra input of energy from the rotating
BH makes the funnel region very energetic (see the bottom-middle panel
of Fig.~\ref{f.5panels}). At radius $r=250$ the energy flux in the
funnel and optically thick outflow is dominated by radiative
luminosity equal to $8.4L_{\rm Edd}$. Most of these photons are
carried with the gas and ultimately will escape when they reach the
photosphere, which for most angles is outside the computational
domain. However, the radiative luminosity may decrease if radiation
keeps transferring its momentum to the gas. The kinetic component of
the luminosity is significant already at this radius ($r=250$) ---
roughly $1.5L_{\rm Edd}$ is carried in the form of outflow kinetic
energy.

The remaining three-dimensional run (\texttt{r020}), which differs
from the fiducial run (\texttt{r001}) in the mass of the BH ($1000\Msun$ instead of
$10\Msun$) shows very similar properties in all respects.

\begin{table*}
\centering
\caption{Luminosities}
\label{t.luminosities}

\begin{tabular}{lC{0.8cm}C{0.8cm}C{0.8cm}C{0.8cm}C{0.8cm}C{0.8cm}C{0.8cm}C{0.8cm}C{0.8cm}C{0.8cm}}
\hline
\hline
&  \multicolumn{2}{c}{\texttt{r001}} & \multicolumn{2}{c}{\texttt{r003}} & \multicolumn{2}{c}{\texttt{r011}} & \multicolumn{2}{c}{\texttt{r020}} & \multicolumn{2}{c}{\texttt{d300} (2D)}\\
\hline
$\dot M /\Medd$ & \multicolumn{2}{c}{10.0} & \multicolumn{2}{c}{175.8} & \multicolumn{2}{c}{17.6} & \multicolumn{2}{c}{10.1} & \multicolumn{2}{c}{8.9} \\
$L_{\rm total}/L_{\rm Edd}$      & \multicolumn{2}{c}{5.26} & \multicolumn{2}{c}{83.27} & \multicolumn{2}{c}{14.27} & \multicolumn{2}{c}{5.14} & \multicolumn{2}{c}{5.31} \\
$\eta_{\rm total}$      & \multicolumn{2}{c}{0.030} & \multicolumn{2}{c}{0.027} & \multicolumn{2}{c}{0.084} & \multicolumn{2}{c}{0.029} & \multicolumn{2}{c}{0.034} \\
\hline
\hline
$r_{\rm lum}$                        & 50     & 250 & 50 & 250 & 50 &
250 & 50 & 250 & 50 & 250 \\                      
\hline
$L_{\rm rad,\,thin}/L_{\rm Edd}$ & 0.30 & 0.95 & 0.0 & 0.0 & 0.0 & 2.13 & 0.20 & 0.85 & 0.64 & 2.14 \\
$L_{\rm kin,\,thin}/L_{\rm Edd}$ & 0.00 & 0.15 & 0.0 & 0.0 & 0.0 & 0.16 & 0.00 & 0.15 & 0.01 & 0.23 \\
$L_{\rm rad,\,out.}/L_{\rm Edd}$ & 1.13 & 1.57 & 0.78 & 3.47 & 6.70 & 8.39 & 1.11 & 1.60 & 1.86 & 2.58 \\
$L_{\rm kin,\,out}/L_{\rm Edd}$ & 0.17 & 0.35 & 0.50 & 4.82 & 1.41 & 1.54 & 0.21 & 0.39 & 0.31 & 0.41 \\
\hline
\hline
\multicolumn{11}{l}{$\dot M$ - average accretion rate}\\
\multicolumn{11}{l}{$L_{\rm total}/L_{\rm Edd}$ - total (conserved) luminosity of
  the system including the flux of binding energy}\\
\multicolumn{11}{l}{$\eta_{\rm total}=L_{\rm total}/\dot M c^2$ -
  efficiency of accretion in total luminosity}\\
\multicolumn{11}{l}{$L_{\rm rad,\,thin}$, $L_{\rm kin,\,thin}$ -
 radiative and kinetic luminosities in the optically
  thin polar region integrated at radius $r_{\rm lum}$}\\
\multicolumn{11}{l}{$L_{\rm rad,\,out}$, $L_{\rm kin,\,out}$ -
  radiative and kinetic luminosities in the region of
  unbound gas ($Be>0$) integrated at radius $r_{\rm lum}$ }\\
\end{tabular}
\end{table*}

\subsection{Angular distribution of energy fluxes}
\label{s.angular}

\begin{figure}
\includegraphics[width=1.\columnwidth]{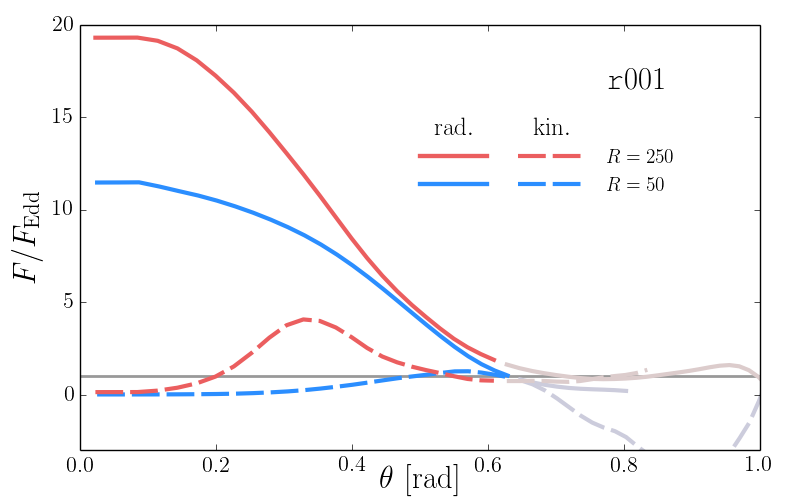}\vspace{-.5cm}
\includegraphics[width=1.\columnwidth]{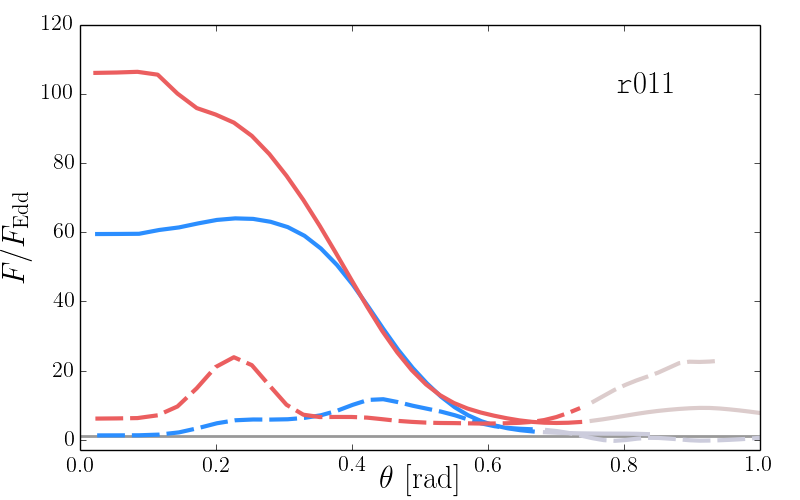}\vspace{-.5cm}
\includegraphics[width=1.\columnwidth]{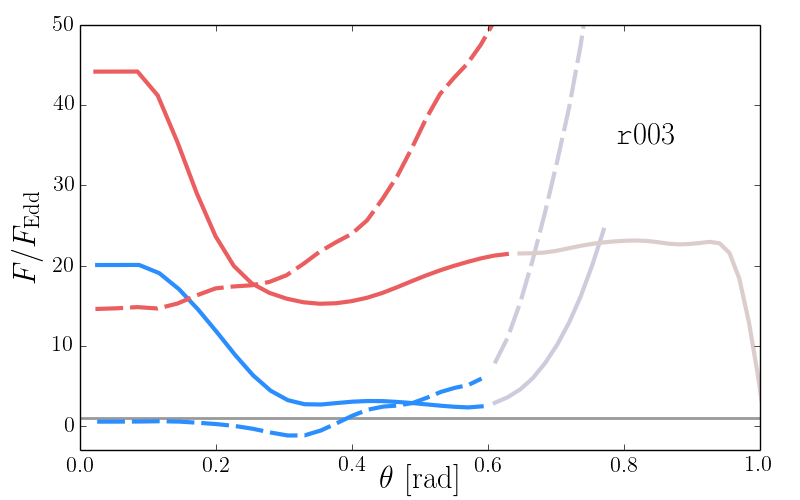}
\caption{Average energy fluxes in simulations \texttt{r001} (top
  panel), \texttt{r003} (middle), and \texttt{r011} (bottom) as a
  function of the polar angle $\theta$. Blue and red lines correspond
  to fluxes measured at $r=50$ and $r=250$, respectively. Solid lines
  show the radiative flux and dashed lines show the flux of kinetic
  energy. Bright line segments are within the outflow region ($Be>0$),
  while the dull shaded segments are outside this region.}
\label{f.fluxvsth}
\end{figure}

Figure~\ref{f.fluxvsth} shows the angular distribution of the (time-
and azimuth-averaged and symmetrized) 
radiative and kinetic fluxes of energy in the funnel/outflow region for three simulations (top to bottom):
\texttt{r001}, \texttt{r011}, and \texttt{r003}. Blue and red lines
correspond to fluxes measured at $r=50$ and $r=250$,
respectively. Solid and dashed lines denote the radiative and kinetic
energies, respectively.

For the fiducial model (top panel, $\dot M=10\Medd$) the energy flux
(as discussed in the previous section) is dominated by
radiation. The angular distribution of radiative flux follows 
roughly a Gaussian with half-maximum width at $\theta=0.35\rm rad$. The maximal flux at
radius $r=50$ is $F\approx12F_{\rm
  Edd}$; it increases to $F\approx19F_{\rm
  Edd}$ at $r=250$. The radiative fluxes decline significantly
with increasing polar angle $\theta$. For an observer at $\theta=0.5\rm
rad$ ($\sim 30^\circ$) the observed flux (and the inferred source
luminosity) is only $\sim 4F_{\rm Edd}$.

The numbers given above are meaningful for the optically thin region but
less so for other angles where the wind is optically thick.
In these regions, there is
still significant interaction between gas and radiation. The
radiation is likely to accelerate the gas further and the radiative flux is expected
to decrease
towards the Eddington limit. To study this effect quantitatively one would
have to perform simulations in a much bigger box and for a much longer time.

At radius $r=50$ there is hardly any kinetic luminosity in the polar
region of the 
fiducial run (\texttt{r001}). Only further from the BH is the gas is accelerated.
In contrast to the radiative flux, the kinetic energy flux is
not concentrated at the polar axis but rather in a shell around
$\theta=0.35\rm rad$, similar to the jet/wind boundary discussed in
\cite{sadowski+outflows}. For the fiducial model accreting at
$10\Medd$ the maximal kinetic flux at that inclination equals $\sim
4F_{\rm Edd}$.

The angular profiles for the run with a rotating BH (\texttt{r011})
look qualitatively similar to the profiles of the fiducial
run. However, the magnitudes of the fluxes are much higher, reflecting
the increased efficiency of accretion due to energy being extracted from
the BH. The maximal radiative and kinetic fluxes at the axis at $r=250$
exceed $100F_{\rm Edd}$ and $20F_{\rm Edd}$, respectively.

A larger mass accretion rate changes the picture significantly. The third
panel of Fig.~\ref{f.fluxvsth} shows the angular profiles of fluxes
for run \texttt{r003}. The radiative flux is still beamed at the axis,
and the maximal flux exceeds $40F_{\rm Edd}$ at $r=250$. The kinetic
energy flux has a different shape than it used to. It is no longer
concentrated in a shell but increases with angle,
reflecting the fact that the wind carries a significant amount of
energy over a wide solid angle. However, even at the axis, the kinetic energy
flux is much higher than in the fiducial model --- it now exceeds $10F_{\rm
  Edd}$ at $r=250$. The kinetic flux in the funnel is a result of radiative energy
being converted into kinetic energy within the optically thick region
\citep{sadowski+radjets}.

The angular distribution of energy fluxes discussed above should be
considered only approximate due to the limitations of the M1 closure
scheme adopted in this work. Most importantly, we evolve only the
first moments of the radiation field instead of evolving specific
intensities directly as in \cite{jiang+14a} and
\cite{ohsugatakahashi-15}. The radiation observed by a distant
observer should, in principle, be calculated as an integral of the
specific intensity pointing towards the observer over the whole
accretion disk. The local radial flux gives only an approximation of
this quantity. Furthermore, the M1 closure is known to have
difficulties in treating the region closest to the polar axis. We have
substantially mitigated this problem by including an extra radiative viscosity
\citep{sadowski+dynamo}, but the coefficients involved had to be
chosen somewhat arbitrarily.

\section{Photon trapping}
\label{s.trapping}

At the time the thin disk theory was established \citep{ss73,novikovthorne73}, it was
commonly understood that the standard radiation pressure dominated
disk can extend up to and above the critical Eddington accretion rate (equation~\ref{e.medd}). Once the Eddington limit is
exceeded, it was predicted that the most energetic, inner
region would attempt to produce a luminosity exceeding the Eddington value and
this would produce a radiatively driven outflow inside the so called
spherization radius. At the same time the wind would modify the accretion rate
on the BH.

This picture did not take an important effect into account. When the
accretion flow is very optically thick,
photons do not have enough time to
diffuse vertically to the disk photosphere before they are dragged radially inward
by the accreting gas and advected
into the BH. This effect was described in a simple spherical context by
\cite{begelman-78} and included for the first time in a full-fledged accretion model
by \cite{abra88} who constructed the so-called slim disk model. These authors
predicted that the radiative luminosity would no longer scale with the
accretion rate above the critical rate. Also, in priniciple, the slim disk state would prevent
the spherization phenomenon.

As we have shown in the previous section, and as we will explain in greater detail here, the
truth is in between --- there is a region, but only near the axis, where locally flux is
significantly super-Eddington and where radiation drives gas out of
the disk, but at the same time photons are trapped and advected
towards the BH in the inflow
region near the equatorial plane.

In this section we try to answer the question: where is photon
trapping effective or, in other words, where is the trapping radius, the
border between the radiatively inefficient (slim disk) and
radiatively efficient (thin disk) regimes?

\subsection{One-dimensional, luminosity-based trapping radius}

It is relatively straightforward to define the trapping radius
in a spherically symmetric flow \citep{begelman-78}. Assuming a
Bondi-like flow, we can compare the local outward
radiative diffusion velocity to the local gas
inflow velocity and thereby estmate
\be
\label{e.Rtrap.bondi}
R_{\rm trap,\,Bondi}\approx \dot M/\Medd.
\ee
This approach gives only an upper limit on the location of the trapping
radius in a real accretion flow; accretion disks are not spherically symmetric but have low
density polar regions where
radiation can more easily escape, and also the inflow velocity near the
equatorial plane is significantly lower than the free-fall velocity
assumed in the Bondi model \citep{sadowski+koral2}.

The simplest way of estimating the location of the trapping radius
numerically is to look at the net flux of radiation, e.g., the dashed
lines in Fig.~\ref{f.radfluxall}. Negative values mean that more
photons were moving inward (mostly because they are dragged by
optically thick gas) than outward. The radius at which the radiative
luminosity goes to zero could then be defined as the trapping
radius. This approach is simple and provides a single effective
trapping radius. It does not, however, account for the non-uniform
structure of the disk -- photons easily leave the system at the axis
and are more effectively trapped near the disk mid-plane. The present
simple definition weights both effects and provides some kind of an
average. By this definition, for simulations \texttt{r001} and
\texttt{r020}, the effective trapping radius is located around
$r=35$. Model \texttt{r003}, which has a significantly larger
accretion rate, has almost the same trapping radius,
$r\approx40$. Model \texttt{r011}, with a rotating BH, produces much
more powerful radiation flux along the axis, and correspondingly the
effective trapping radius is much closer to the BH ($r\approx 10$,
although the trapping in the bulk of the disk is as effective as in
the other cases). A caveat: The values given above for the simulations
with non-rotating BHs should be taken with caution, because the bulk
of the disk is outside the inflow/equlibrium region at these radii.

\subsection{Two-dimensional trapping - importance of the diffusive
  flux}

Because of the limitations of the definition given in the previous
section, we discuss here a different approach which, in particular,
seeks to account for local properties of the flow.

\cite{jiang+14b} calculated energy-weighted, vertically integrated,
radial and vertical velocities of radiation transport. By comparing the two one can distinguish whether
more energy flows up (away from the midplane) or inward (towards the
BH).
not adequate to say where the photon trapping is effective because it
does not gives no importance to diffusion. The vertical flow of
radiation could be the result of advection (in a wind) or turbulence \cite{jiang+14b}.
Thus, while it is a good way of comparing vertical and
radial energy transfer, it provides no information on
the transport mechanism.

Below we attempt to quantify photon trapping by calculating the
fraction of the total radiative flux that comes from diffusive
transfer. To estimate the diffusive flux we use the moment equations
(\ref{eq.rmunucons}), assuming $\partial_t=0$ and a diagonal form of the
radiative stress energy tensor in the fluid frame ($\widehat
R^{ii}=1/3\widehat R^{tt}$), to obtain \be
\label{e.Fdiff}
F^i_{\rm diff}=\frac{1}{3\avg{\kappa \rho}}\frac{d}{dx^i}\avg{\widehat E},
\ee
which is the standard diffusive flux formula
\citep{rybicki-book}. This expression should be used only inside the
optically thick parts of the disk where the Rosseland approximation is
satisfied.

\begin{figure}
\includegraphics[width=1.\columnwidth]{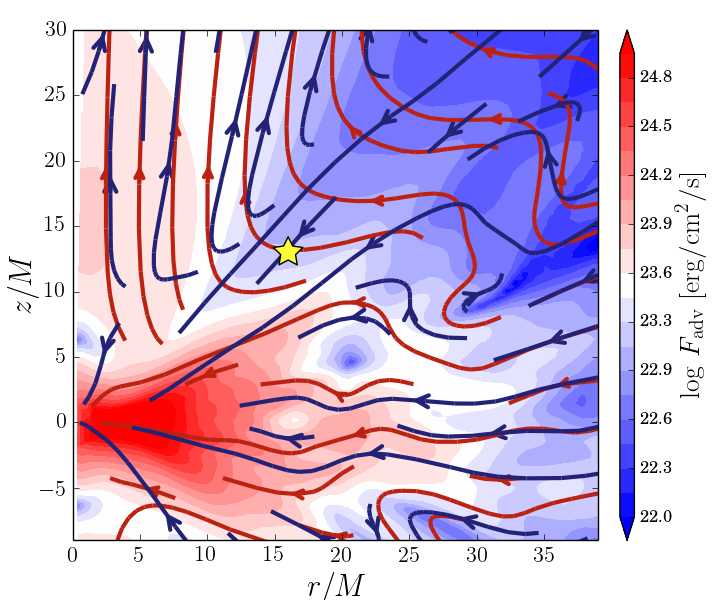}\vspace{-.4cm}
\includegraphics[width=1.\columnwidth]{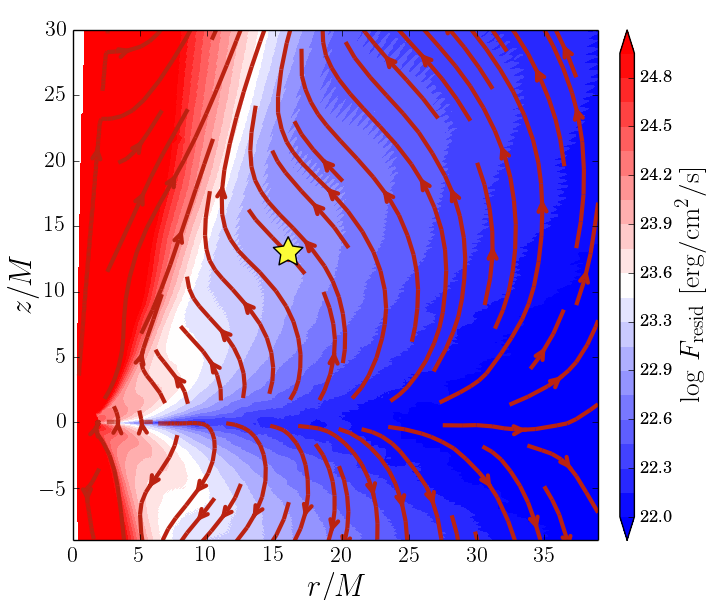}\vspace{-.4cm}
\includegraphics[width=1.\columnwidth]{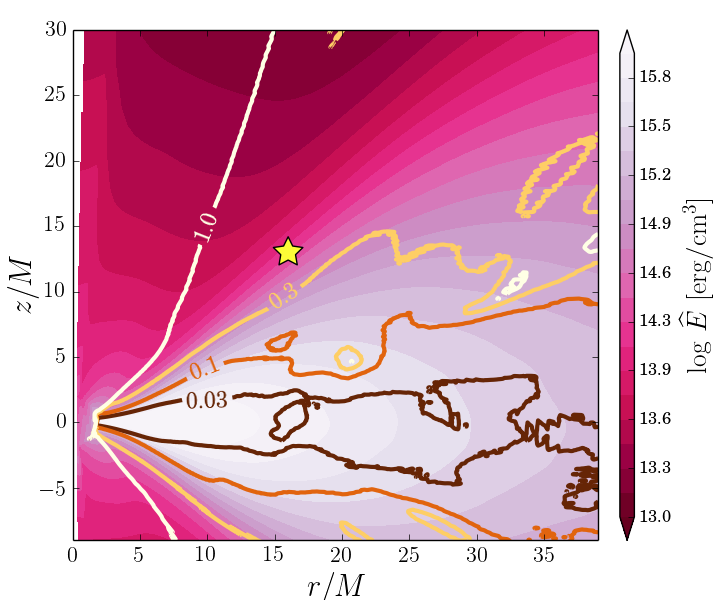}
\caption{Top panel: Magnitude and streamlines of the total radiative
  flux $F^i_{\rm rad}=-R^i_t$.  Middle panel: Magnitude and direction of the radiative
  diffusive flux $F^i_{\rm diff}$ (equation~\ref{e.Fdiff}). Bottom
  panel: Magnitude of the comoving frame radiative energy density
  (colors) and the ratio of the diffusive radiative flux to the total
  radiative flux ($F^i_{\rm diff}/F^i_{\rm rad}$). All panels
  correspond to simulation $\texttt{r001}$ ($10\medd$).  The yellow
  stars denote the location where we studied in detail the vertical
  fluxes plotted in Fig.~\ref{f.veldiff}.}
\label{f.phottrap}
\end{figure}

\begin{figure}
\includegraphics[width=1.\columnwidth]{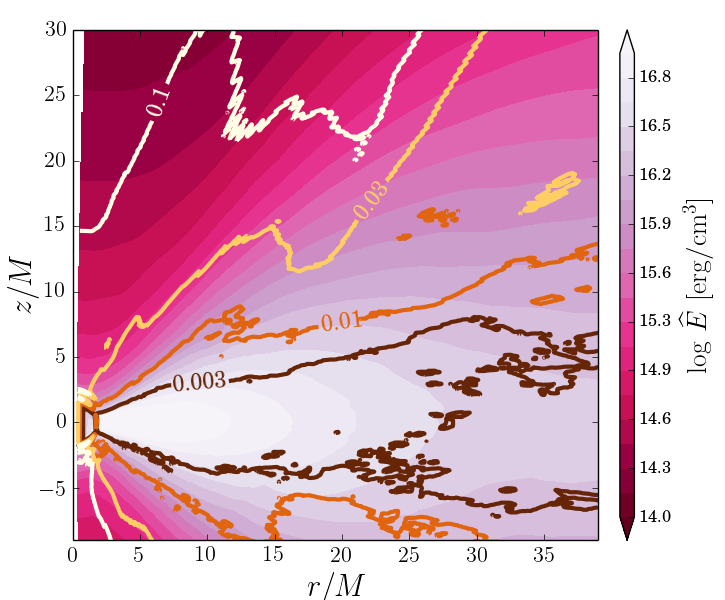}
\caption{Magnitude of the comoving frame radiative energy density
  (colors) and the ratio of the diffusive to advective radiation flux
  ($F^i_{\rm diff}/F^i_{\rm rad}$) in simulation $\texttt{r003}$
  ($175\medd$).}
\label{f.phottrap_r003}
\end{figure}

The top panel of Fig.~\ref{f.phottrap} shows the magnitude
  (colors) and
direction (red streamlines) of the total radiative flux, $F^i_{\rm tot}=R^i_t$,
for the fiducial model \texttt{r001} which accretes
at $10\medd$. Deep in the disk the radiation flows towards the
BH, while the
polar region is filled with
optically thin radiation escaping along the axis. There is a
transition region at intermediate
angles, $\theta\approx 45^\circ$, where radiation relatively smoothly
switches from flowing radially inward to flowing outward.

Blue streamlines in the same panel show the average gas velocity. If
all the radiative flux was coming from photons advected with the gas,
the total radiation flux vectors would follow everywhere the direction
of the gas velocity. The agreement between the two sets of streamlines
is very deep inside the disk, where both gas and
radiation flow inward. This is where gas is most optically thick and
where one expects efficient photon trapping. The streamlines agree again
in the polar region, but this is not because of photon trapping. The low
optical depth allows radiation to flow independently of the gas. The
funnel geometry, however, makes both gas and radiation to flow upward, and the locally
super-Eddington flux accelerates gas in its direction. In the
intermediate region between the funnel and the disk interior, the gas
velocity and radiation flux vectors do not point in the same
direction. As mentioned previously, gas flows on average radially inward
in the disk and only close to the surface is it blown away, causing the velocity streamlines
to turn rapidly outward. Radiation flux, on the other hand,
changes direction rather smoothly. This difference between gas and radiation
streamlins shows that there must
be an additional component of the total flux besides the advective
photon transport. This is of course diffusive transport.

The middle panel in Fig.~\ref{f.phottrap} shows on the same axes and
with the same color scale the diffusive flux estimated according to
equation (\ref{e.Fdiff}). As expected, the diffusive flux follows the
gradient of radiative energy density, i.e., radiation diffuses in the
vertical direction away from the equatorial plane. Then, it turns
smoothly towards the axis and enters the funnel region. IN the funnel
itself (the red region), the estimate of the diffusive flux must be
disregarded since gas there is optically thin and radiation is free
streaming instead of diffusing. However, the diffusive flux estimates are
trustworthy in the disk interior and the transition
region between the disk and the funnel. 

\begin{figure}
\includegraphics[width=1.\columnwidth]{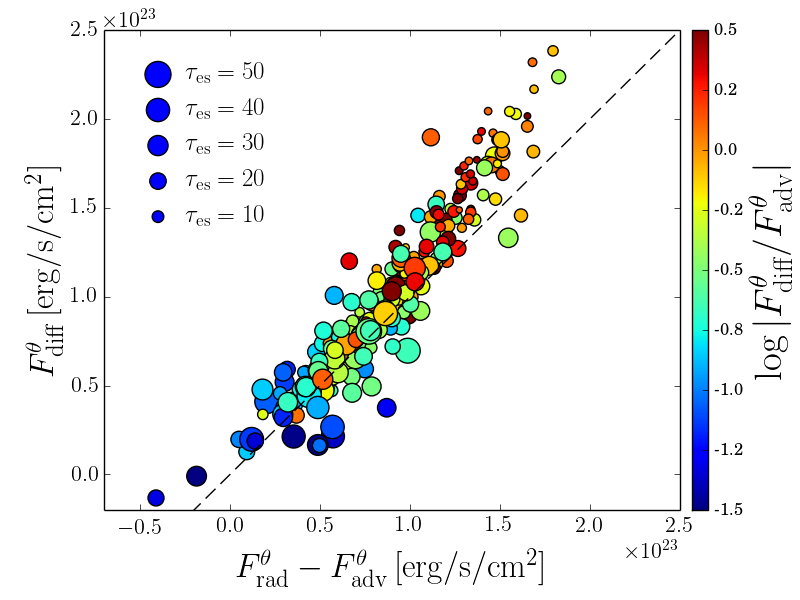}
\caption{Estimated polar diffusive flux of radiation
  (equation~\ref{e.Fdiff2}) versus the excess of the total flux over
  the local advective flux of radiation
  (equation~\ref{e.Fsub}). Colors denote the local optical depth over
  a distance of one gravitational radius. The dashed line shows where
  the two fluxes agree, i.e., when the excess can be explained
  entirely by diffusive transport. The data are from simulation
  $\texttt{r001}$ and the points range from $t=7,500\div20,000$ with a
  cadence of $dt=50$.}
\label{f.veldiff}
\end{figure}

We now check whether the diffusive flux is, in fact, responsible for
the deviation between the total radiative flux streamlines and the
advected radiation flux streamlines (which would be aligned with the
gas velocity).  For this purpose we choose a location in the region of
largest deviation, $(r,\theta)=(20,51^\circ)$ (denoted by the yellow
stars in Fig.~\ref{f.phottrap}), and study local fluxes of radiation
over time. Fig.~\ref{f.veldiff} presents the estimated orthonormal
polar component of the diffusive flux, calculated as, \be
\label{e.Fdiff2}
F^{\hat\theta}_{\rm diff}=-\frac{1}{3\kappa \rho}\frac{d}{rd\theta}\widehat E,
\ee
as a function of the difference between the polar components of the
total flux and the estimated advective flux, 
\be
\label{e.Fsub}
F^{\hat\theta}_{\rm rad}-F^{\hat\theta}_{\rm adv}=\left(R^\theta_t
+\frac{4}{3}\widehat E u^\theta \right)r.  \ee The signs have been
chosen such that the positive values correspond to fluxes pointed out
of the disk, i.e., towards the axis. The sizes of markers measure the local
optical depth for scattering as estimated by
$\tau_{\rm es}=\kappa_{\rm es}\rho R_G$. The dashed line
denotes $F^{\hat\theta}_{\rm diff}=F^{\hat\theta}_{\rm
  rad}-F^{\hat\theta}_{\rm adv}$. The plotted points cover the time period
$t=7,500\div20,000$ with cadence of $\Delta t=50$.

The majority of points in Fig.~\ref{f.veldiff} cluster around the
dashed line, showing that the excess of radiative flux over advective
flux has exactly the analytically predicted magnitude, i.e., the
excess of polar radiative flux is due to diffusion. The agreement is
no longer perfect at the lowest optical depths. This is natural since,
when the optical depth is low, radiation decouples from the gas and
the diffusive approximation no longer valid. In fact, this is where we
expect the actual flux to lie below the analytically estimated diffusive
flux, as is indeed seen in the plot.

The colors of the markers in Fig.~\ref{f.veldiff} denote the ratio of
magnitudes of instantenous diffusive and advective fluxes. Despite the
fact that the time-averaged advective flux has hardly any polar
component, instantenous advective flux in $\theta$ direction can be
more than 30 times stronger than the corresponding diffusive
flux. However, because of turbulent motion of gas, it averages to a
value significantly smaller than the diffusive flux. Therefore, it is
the diffusive flux which dominates the average net radiation flux
towards the axis near the surface of the disk.

The bottom panel in Fig.~\ref{f.phottrap} shows the ratio of the magnitude of the diffusive
flux over the total flux plotted with contours on top of the
distribution of fluid frame radiative energy density, $\widehat
E$. Deep in the disk, advection dominates strongly over
diffusion. The magnitude of the former flux is $\gtrsim 30$ times
larger than the magnitude of the diffusive flux. This ratio is lower
closer to the disk surface, where the density drops and the gas
is no longer able to drag photons efficiently. In the transition
region discussed above, the advective and diffusive
fluxes have comparable magnitudes, with the diffusive flux dominating
the polar component. 

Figure~\ref{f.phottrap_r003} shows the same ratio of the diffusive to
advective fluxes, but for simulation $\texttt{r003}$ which has almost $20$ times
larger accretion rate. The larger accretion rate implies higher gas
density and optical depth, and, as a result, more effective photon
trapping. This is indeed the case. The advective radiative flux is now
$\gtrsim 300$ times larger than diffusive in the disk near the equatorial
plane.

The properties described above show that for moderately super-critical
accretion rates, $\sim 10\dot M_{\rm Edd}$, photon transport deep
inside the disk is dominated by radial advection. However, diffusive
transport of energy becomes important further from the equatorial
plane, near the transition between the disk and the funnel, where the
gradient of the energy density is large, and the optical depth of the gas
is no longer huge. Radiation in the optically thin funnel follows the
axis and is decoupled from gas. Because of significant photon trapping
deep in the disk, only photons generated close to the surface can
escape the disk and join the funnel region. This explains the relatively low
radiative luminosities of our simulated disks and is in agreement with the slim
disk model \citep[e.g.,][]{abra88,sadowski.phd}. For higher accretion
rates, the region of dominant photon trapping extends further out, not
only near the equatorial plane, but also in the (now optically thick)
funnel. Likewise, one expects that, for lower accretion rates, photon
trapping will become less and less effective and the radiative
efficiency will approach the thin disk value.

\subsection{Turbulent transport}{

So far we have shown that the diffusive transport of radiation is
important near the surface of the disk, and that the radial photon advection
dominates over diffusion deep inside the disk. There is potentially
one more way of radiation transfer - turbulent
transport pointed out by \cite{jiang+14b}. It is effective if
radiative energy is transported without transporting mass, similar to
what happens in convection. Such behavior may result, e.g., from
magnetic buoyancy. To assess the importance of this effect it is
enough to compare the density- and radiative energy-weighted vertical
velocitities of the gas. If the former is zero and the latter is not, then radiation is
transported by gas without transporting mass, and hence, it is
different in nature from the standard photon trapping which results
from mean motion of gas.

If turbulent transport is effective then lighter gas,
but containing more radiation, moves preferably towards the disk surface. In
Fig.~\ref{f.turbtrans} we plot the polar velocity of gas as a function
of radiation to rest-mass energy density ratio for the same location
as in Fig.~\ref{f.veldiff} (red), and for another point closer to the
equatorial plane, located at $(r,\theta)=(20,80^\circ)$ (blue
markers). The size of the markers denotes the optical depth as it did
in Fig.~\ref{f.veldiff}. We see that there is no correlation between
the relative radiative energy density content and the vertical motion
of the gas, what suggests that the turbulent effect cannot be strong.

From the same sets of points we now calculate the density- and
energy-weighted velocities, defined as,
\bea
\hspace{2cm}\avg{v^\theta}_{\rho}&=&\frac{\Sigma\, \rho v^\theta}{\Sigma\, \rho},\\
\avg{v^\theta}_{\widehat E}&=&\frac{\Sigma\, \widehat E v^\theta}{\Sigma\,\widehat
  E },
\eea
where the sums go through all the points in the set, corresponding to
different moments of time and fixed location. For the point closer to
the disk surface we get $\avg{v^\theta}_{\rho}=0.0022$ and
$\avg{v^\theta}_{\widehat E}=0.0013$. For the other point, located
almost at the equatorial plane, we have, $\avg{v^\theta}_{\rho}=0.0038$ and
$\avg{v^\theta}_{\widehat E}=0.0002$. All the values are positive what
indicates that both the gas and radiation is advected with the gas
\textit{towards} the equatorial plane. However, the magnitudes of
these velocities are at least order of magnitude lower than of the
corresponding radial velocities, what reflects the fact that the
radial motion and advection dominates. One may, however, notice that
the energy-weighted velocities are lower, i.e., they tend to deviate
from the density-weighted velocities towards the
surface. Nevertheless, the magnitude of the turbulent transport of
radiation is not significant.

\begin{figure}
\includegraphics[width=1.05\columnwidth]{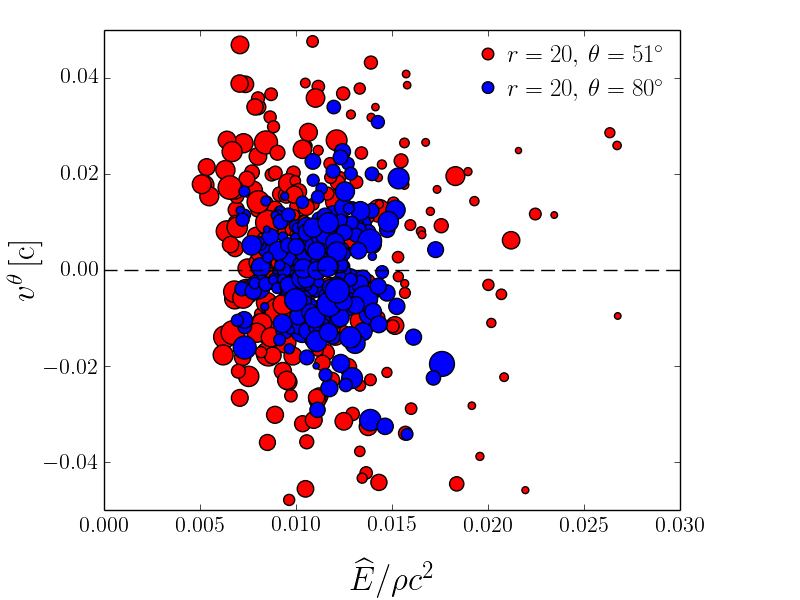}
\caption{Vertical velocity of gas as a function of radiative to
  rest-mass energy ratio for simulation \texttt{r001} and two points,
  located at $(r,\theta)=(20,851^\circ)$ (red) and
  $(r,\theta)=(20,80^\circ)$ (blue markers).}
\label{f.turbtrans}
\end{figure}

\subsection{Advection coefficient}

Defining the efficiency of photon trapping in a non-spherical
accretion flow is a challenging task because of the multiple
dimensions involved. Here we try to calculate the advection
coefficient which estimates the fraction of photons generated at given
radius that are able to escape, the rest being advected to the BH. For
this purpuse, at each radius $r_{\rm box}$, we consider a disk annulus
extending from $r_1=r_{\rm box}-0.5 r_{\rm G}$ to $r_2=r_{\rm box}+0.5
r_{\rm G}$, and limited in $\theta$ to the range
$\theta_\pm=90^\circ\pm37^\circ$ (Fig.~\ref{f.box}). The polar angle range
is chosen to fit the center of the transition region discussed in
the previous section where the radiation flux in the fiducial model is
dominated by its polar component. The annulus covers extends over $2\pi$ in
azimuthal angle. From the time averaged disk properties we extract
luminosities of the radiative flux that crosses each surface of the annulus. For the
radial sides, we compute \be L_{r,(1,2)}=\int_0^{2\pi}
\int_{\theta_-}^{\theta_+}R^r_t\sqrt{-g}\,d\theta d\phi, \ee where the
integration takes place either at the inner edge $r=r_1$ or at the
outer edge $r=r_2$. The fluxes crossing the top and bottom surfaces are
similarly integrated to give, \be L_{\theta}=2\times\int_0^{2\pi}
\int_{r_1}^{r_2}R^\theta_t\sqrt{-g}\,dr d\phi, \ee where the
integration is done at fixed polar angles, $\theta_\pm$.

\begin{figure}
\includegraphics[width=1.\columnwidth]{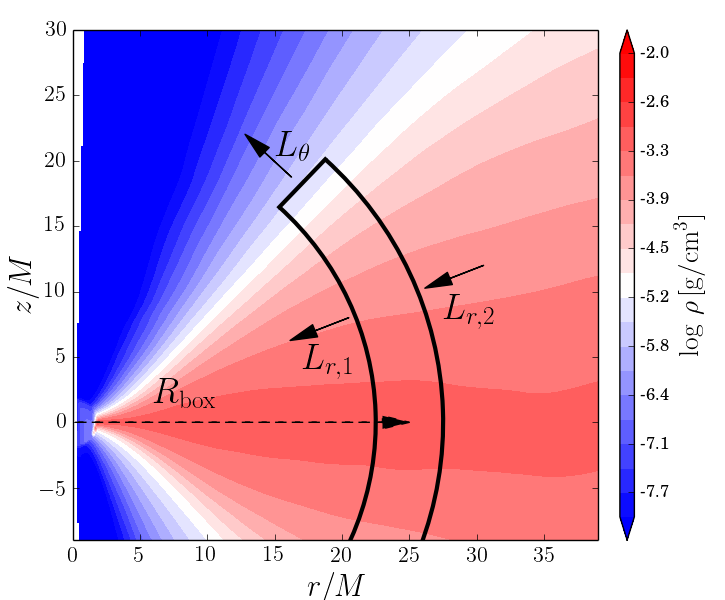}
\caption{Shows the box and the definitions of the luminosities used in
  equation~(\ref{e.qadv}) to estimate the advective factor $q_{\rm
    adv}=1-L_\theta / (L_{r,1}-L_{r,2}+L_{\theta})$.}
\label{f.box}
\end{figure}

The integrated flux $L_{r,1}$ incoming through the outer radial sufrace tells
how much radiation is advected into the volume. The corresponding
luminosity $L_{r,2}$ crossing the inner surface tells how much radiation
is advected out of the same volume. Meanwhile, the total radiation generated within
the annulus is $L_{r,1} + L_\theta - L_{r,2}$. Thus $L_{r,1} - L_{r,2}$ divided
by the latter quantity measures the fraction of energy that is advected radially,
whereas $L_\theta$ over the same quantity measures what fraction of the 
radiative energy escapes through the top and bottom surfaces. This then motivates
the following
definition of the advection coefficient, $q_{\rm adv}$,
\be
q_{\rm adv}=\frac{L_{r,1} - L_{r,2}}{L_{r,1}-L_{r,2}+L_{\theta}}.
\label{e.qadv}
\ee In the limit of a radiatively efficient disk we would have
$L_{r,1}=L_{r,2}\approx 0$ and $q_{\rm adv}\approx0$. In the opposite
limit of an advection-dominated flow we expect $L_{\theta}=0$ and
$q_{\rm adv}\approx1$. Tehrefore, it is natural to define the
effective trapping radius as the location where $q_{\rm adv}=1/2$, i.e.,
where half the radiation generated at that radius manages to
escape and half is advected to the BH.

Figure~\ref{f.qadv} shows the above advection coefficient $q_{\rm
  adv}$ as a function of radius for the fiducial model \texttt{r001}
and the large accretion rate model \texttt{r003} (the chosen vertical
size of the box corresponds to the region of purely polar radiative
flux only for the former). For the fiducial simulation, the advection
factor increases towards the BH, as expected because of increasing
inflow velocity. At radius $r=10$ we find $q_{\rm adv}=0.8$ which
means that only $\sim 20\%$ of photons generated at that radius manage
to escape and enter the funnel. The coefficient drops down to $q_{\rm
  adv}=0.55$ at the edge of the inflow/outflow equilibrium region
($r=25$). The profile suggests that the effective photon trapping
radius, defined by $q_{\rm adv}=1/2$, is probably around $r\approx 35$
(the point plotted at $r=30$ is outside the region of inflow
equilibrium and is a little suspect). 

The efficiency of advection for the simulation with higher accretion
rate (\texttt{r003}) is significantly larger because of the larger
optical depth. Even at radius $r=25$ only $\sim 15\%$ of photons
escape the disk. In the innermost regions of this model, $q_{\rm adv}$
exceeds $1$, reflecting the fact that the top-bottom luminosities are
negative, i.e., photons are brought into the box, and no radiation
escapes.

In our analysis in this subsection we focused solely on the radiative
energy flux. In general, one should include also the flux of
mechanical energy because dissipation may result in kinetic energy
leaving the box. Including this component hardly affects the values of
the advection coefficient calculated above.

\begin{figure}
\includegraphics[width=1.\columnwidth]{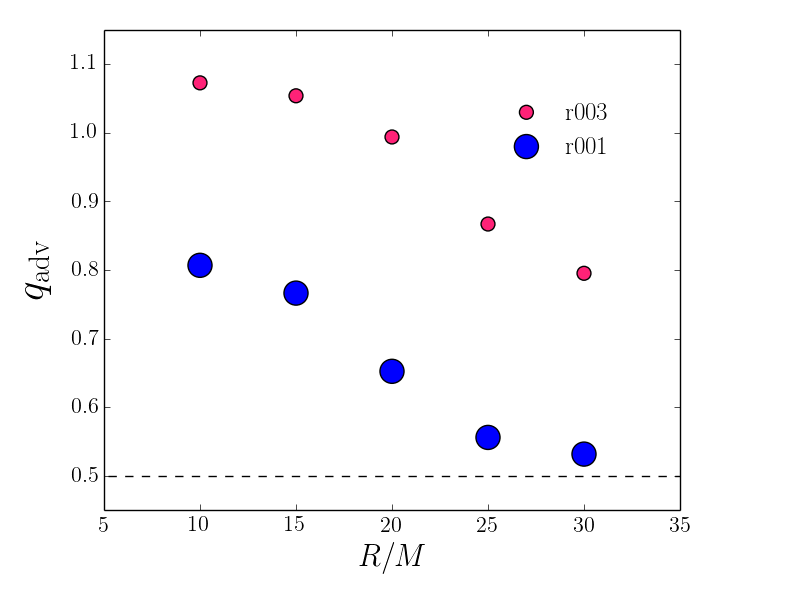}
\caption{The advection coefficient, $q_{\rm adv}$
    (equation~\ref{e.qadv}) as a function of radius for simulations
    \texttt{r001} (blue) and \texttt{r003} (red circles).}
\label{f.qadv}
\end{figure}

\section{Variability}
\label{s.variability}

Radiation coming from accretion disks is known to be highly
variable \citep[e.g.,][]{done+07}. In case of galactic BH binaries, this
variability takes place on short timescales (the horizon crossing time
for a $10\Msun$ BH is $GM/c^3\approx 5\times 10^{-5}\rm s$). The variability is
strongest in the hard state, and weakest in the thermal state.
However, even in the latter, the power spectrum is far from
featureless. 

Studying variability is a powerful tool in understanding accretion
flows. The characteristic frequencies tell us where the modulated radiation
come from. Features in the power-spectrum, e.g., quasi-periodic
oscillations or breaks, can manifest more subtle properties of the
disks \citep[e.g.][]{ingram+07,wellons+14}. Modeling the variability
and its power spectrum has so far been limited 
mostly to
analytical models. However, the dynamics of the gas and the properties of the
radiation field are complicated and highly non-stationary, so the
analytical approach is limited, and ultimately we should model
variability using time-dependent three-dimensional simulations.

However, this approach is not straightforward. Numerical modeling of
radiation in MHD codes is limited by a number of
factors. First of all, 
radiation is solved for in the grey approximation and some arbitrary
(usually blackbody) shape is often assumed for the spectrum. Secondly,
general relativistic effects are rarely included. Moreover,
Comptonization is either neglected or treated in a crude way. Last but not
least, various approximations for radiation closure are adopted (with
the exception of direct radiation transfer solvers operating on a fixed
grid of angles, as implemented recently by \cite{jiang+14a} and
\cite{ohsugatakahashi-15}). Simplistic closure schemes may limit the information available
for calculating the visible spectra.

The most reasonable way to proceed is to take the global,
time-dependent output of a disk simulation and postprocess it with a
sophisticated radiation (and only radiation) solver which will not be
as limited as full radiation-MHD simulations.  Such codes, which
solving the frequency-dependent radiative transfer equation and
account for relativistic effects and Comptonization have recently been
developed \citep[e.g.][]{zhu+15} and are expected to be soon available
for spectral modeling.

In the meantime, we attempt to directly estimate the variability of
light curves from the simulations with \koral described in this
paper. Because of the limitations mentioned above, in particular the
fact that we evolve only the first moments of the radiation (M1
closure), our approach is expected to provide only a rough qualitative
understanding of the temporal properties of radiation coming from
super-Eddington accretion flows.

\subsection{Light curves}

In principle, the observer is located at infinity,
i.e, $r\gg r_{\rm out}$. The correct way of measuring the radiation reaching a
distant observer is to integrate the specific intensities pointing
towards the observer at the outer edge of the computational box. However, because
of the limited range of inflow/outflow equilibrium near the equatorial
plane, we are limited only to studying light escaping in the polar
region near the axis, where the disk solution has converged to large
enough radii. Moreover, the size of our computational box is
obviously limited and we cannot measure the radiative fluxes at radii
larger than $r \gtrsim r_{\rm out}/2 = 500$. Even if the duration of
simulation was infinite, the photosphere would be located beyond the
domain boundary except in the polar region, and even there we resolve
the photosphere only at the lowest accretion rates.

Because
of the limitations mentioned above, we decided to estimate the light
curve by looking at the local flux of radiation at some location
inside the polar region.
 To be sure that the
light curve measured in this way is close to the light curve
seen from infinity, we insist that the
radiation is already decoupled from the gas;
otherwise, continued interaction with (and acceleration of) the gas
could drastically change the properties of the radiation that finally escapes.
Once this condition is satisfied, and provided there is little radiation coming from
other regions of the disk towards the chosen line of sight, the
shape of the light curve should be
independent of the radius it is measured at, only shifted in time by
the light crossing time. Below we test if this criterion is satisfied
for the locally measured fluxes in simulation \texttt{r001}.

\begin{figure}
\includegraphics[width=1.0\columnwidth]{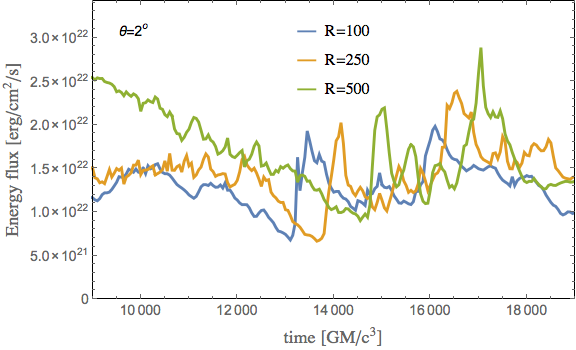}\vspace{-.75cm}
\includegraphics[width=1.0\columnwidth]{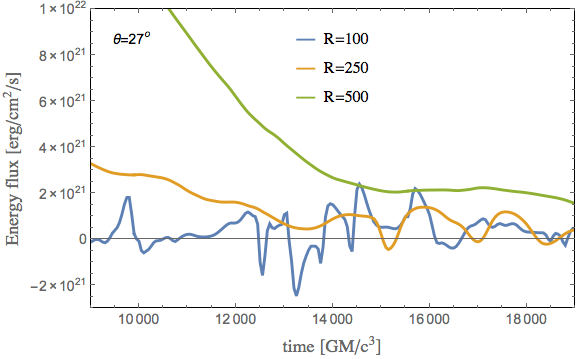}
\caption{Variability of the radiative flux in model \texttt{r001}
  measured at polar angle $\theta=2^\circ$ (top panel) and $27^\circ$
  (bottom panel) and at three radii: $r=100$, $250$, and
  $500$.}
\label{f.fluxvst_r020}
\end{figure}

\begin{table}
\caption{Fractional variability of radiative flux in model \texttt{r001}}
\centering
\begin{tabular}{lcc}
\hline
\hline
   &  $\theta=2^\circ$   &   $\theta=27^\circ$ \\
\hline
$r=100$ &   0.20 & 1.96  \\
$r=250$ &   0.23 & 0.64  \\
$r=500$ &   0.23 & 0.58\\
\hline
\hline
\end{tabular}
\label{t.fracvar}
\end{table}

To extract the light curves we choose an arbitrary slice through the poloidal
plane, define the inclination angle, and choose a radius where we
measure the energy fluxes. Then we
take the radial component of the lab-frame radiative flux, $R^r_t$, and plot it
as a function of time. Fig.~\ref{f.fluxvst_r020} shows such time
series extracted at two polar angles ($\theta=2^\circ$ and $27^\circ$,
top and bottom panels, respectively), and at three radii ($r=100$,
$250$, and $500$). Only the second half of the simulation, when the
accretion has settled down to a quasi-steady state (compare
Fig.~\ref{f.mdots}), is shown. The fluxes measured at $r=100$ and
$r=500$ were scaled to account for geometrical expansion to match the
magnitude of the flux at $r=250$.

The first panel corresponds to radiation escaping along the
axis. For model \texttt{r001}, gas is optically thin on the axis at all
radii so one may expect the measured fluxes not to depend on
the radius where the measument is performed. This
is indeed the case. The profiles resemble
each other to a good accuracy, at least for $t>12000$, with the shift
in time
reflecting the propagation of light in the vertical direction. This gives
some confidence that the light curve we calculate is what a distant observer
might see. It is
interesting, however, that the propagation speed down the funnel to match
the time delay between radii is less then the speed
of light. It equals approximately $0.25c$ which is the characteristic
velocity of the radiation field (the velocity of the radiation rest frame)
inside the funnel. It is determined by the geometry of the funnel and
the effect was discussed close to 40 years ago by \cite{sikora-81}. Basically,
not all the photons go straight up; a significant
fraction go sideways and some even go backward. The characteristic
velocity is in some sense the average radial photon velocity.

The short duration of the light curves ($\sim 5s$ real time for a $10M_\odot$ BH)) do not
allow us to calculate power spectra. Instead, we calculate
the fractional variability through
\be
\label{e.fracvar}
f=\frac{\sigma \left(L(t)\right)}{\avg{ L(t)}}, \ee where $\sigma$
denotes the standard deviation and $\avg{}$ stands for the mean
luminosity. The second column of Table~\ref{t.fracvar} gives the
fractional variability calculated for light curves in the top panel of
Fig.~\ref{f.fluxvst_r020}. Note that the fractional variability does
not change much with radius. In particular, it does not change between
$r=250$ and $r=500$ which again reflects the fact that the radiation
flowing along the axis has established its temporal profile and is not
affected any more by interaction with gas or radiation at larger
radii.

The bottom panel of the same figure presents radiative light curves
measured at the same radii but at a larger inclination angle,
$\theta=27^\circ$, which corresponds to the edge of the funnel region, but
located in the optically thick portion of
the gas. One may expect therefore, that the radiation interacts (is
absorbed or emitted) with the gas up to large radii, even out to the
domain boundary, and that the radiation fluxes measured at different
radii will differ, even after taking the geometrical factor into
account. The radiation flux measured at $r=100$ (blue line) varies
significantly, reaching at moments negative values. This behavior
reflects the motion of the gas to which the radiation is coupled
to. The gas on average moves out, but temporarily can move inward,
dragging photons with it. The light curves become more smooth and the luminosities become
larger in magnitude with increasing radius. These result from
the fact that the gas moves in a more laminar way further out, and 
that radiation originating from a larger volume contributes to the
measured luminosity. As a result, the corresponding fractional variability (third
column of Table~\ref{t.fracvar}) decreases significantly with increasing radius.

Because of the significant coupling with gas, and the fact that we do
not resolve the photosphere (because the computation box is not large
enough box and the duration of the simulation is too short), the light
curves for the larger inclination angle are not reliable. Only the
light curve for radiation leaving the system almost exactly along the
axis is robust. However, even this will not be the case if the
accretion rate is larger since then, even on the axis, the photosphere
will be located at large radii. This is a severe limitation for
studies of variability in
super-critical disks. In contrast, for thin disks,
the photosphere is close to the equatorial plane and one can hope to 
extract useful variability information from simulations.

\section{Axisymmetric simulations}
\label{s.axisymmetric}

Angular momentum in accretion disks is transported by MHD
turbulence, which is driven primarily by the
magnetorotational instability \citep{balbushawley-91}. Turbulent motion in the poloidal plane
leads not only to the exchange of angular momentum, but also to
dissipation of magnetic field. In axisymmetric simulations this
process quickly causes the magnetic field to decay, and at some point,
shortly after the beginning of a simulation, typically after $t \sim5000$,
the accretion stops. In reality, the poloidal magnetic field is
revived by a dynamo process which results from breaking
axisymmetry --- non-axisymmetric perturbation can ``rotate'' toroidal
field into the poloidal plane and thereby regenerate poloidal field. Three dimensional simulations
of turbulent magneto-fluids in shearing boxes (and also in global
models) have shown that the saturated state is characterized by an
average toroidal-to-radial magnetic field ratio $\theta_B\approx
0.25$, and an average magnetic-to-gas pressure ratio
$\beta'=0.1$. There is no way of obtaining such a saturated state in
pure axisymmetric simulations.

On the other hand, the advantages of assuming axisymmetry and
simulating accretion disks in two dimensions are obvious -- such
simulations are cheaper by more than an order of magnitude in terms of
the required computational time. Until recently, however, axisymmetric
simulations were limited to extremely short durations because of the
rapid decay of the magnetic field mentioned above. The situation has
now changed.  \cite{sadowski+dynamo} introduced a mean-field model of
the dynamo effect which mimics the properties of three-dimensional
MRI-driven turbulence but can be applied to axisymmetric
simulations. In this approach, the properties of the magnetic field
are driven towards the prescribed characteristics of the saturated
state, described by parameters $\theta_B$ and $\beta'$. This is
achieved by ``pumping'' new vector potential into the MHD flow,
leading to a correction to the existing poloidal field. The poloidal
field is enhanced in regions where the magnetic field is too weak, and
the toroidal component of the field is damped in regions were the
magnetic pressure exceeds the prescribed saturation value. It has been
shown that such an approach successfully allows for arbitrarily long
simulations and indeed leads to a saturated state similar to that seen
in three-dimensional simulations.

Our aim in this section is to compare the properties of
three-dimensional simulated super-critical, radiative disks with
corresponding two-dimensional axisymmetric simulations. For this
purpuse we simulated an additional model (\texttt{d300}) which was run
in axisymmetry, with $252x234$ grid points in the poloidal plane, and
initialized in an identical manner to the fiducial model \texttt{r001}
(see Table~\ref{t.models}). We used the mean-field dynamo model with
the fiducial saturated state parameters $\theta_B=0.25$ and
$\beta'=0.1$. Model \texttt{d300} was run until $t=190,000$, an order
of magnitude longer than the three-dimensional model \texttt{r001},
yet it required less computer resources by almost an order of
magnitude. This enormous saving is the reason why we feel it is
important to explore the two-dimensional option fully.

The left- and right-most panels in Fig.~\ref{f.5panels} compare the
density distribution on the poloidal plane, gas velocities, 
radiative flux, location of the photosphere, and border of the outflow
region in the three- (\texttt{r001}) and two-dimensional (\texttt{d300}) simulations.
Qualitatively, the two solutions (after averaging over
time and azimuth in the three-dimensional case) agree very well. The
gas shows the same dynamical properties, with outflow taking place
only in the funnel region, the magnitude of the radiation flux is
similar, and the photosphere is located at the same place. The only
noticeable difference is near the disk surface at larger radii
($r\gtrsim 20$) where the two-dimensional run shows more significant
vertical motion. This may be because of the approximate treatment of the
dynamo effect which is constructed to satisfy only the vertically
averaged criteria, without paying too much attention to the vertical
profile of magnetic field properties. It should also be kept in mind that the
three-dimensional model has achieved inflow/outflow equilibrium
only for $r\lesssim 20$. It could be that even the relatively minor differences
in structure between the three- and two-dimensional models would be reduced
once the former is run for a much longer time and reaches equilibrium
out to radii $\sim 100$. Very expensive simulations
would be required to verify if this explanation is correct.

\begin{figure}
\includegraphics[width=1.\columnwidth]{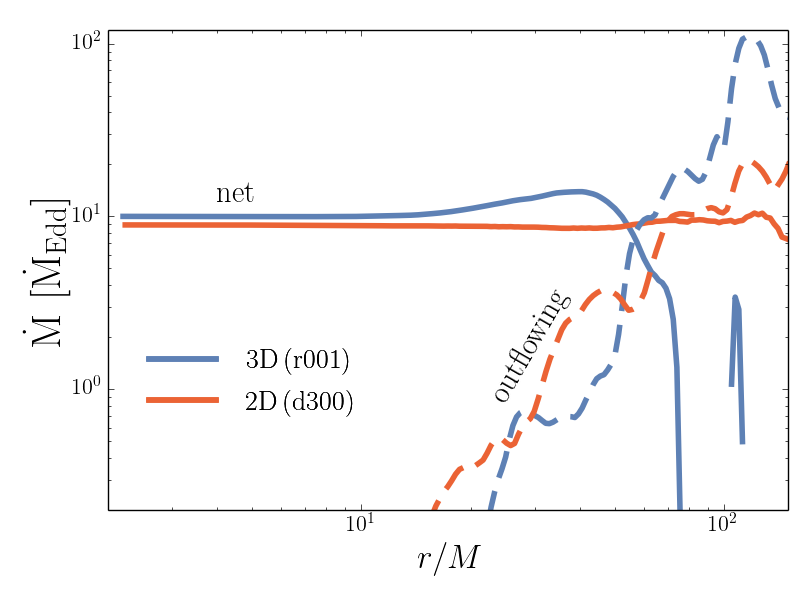}\vspace{-.95cm}
\includegraphics[width=1.\columnwidth]{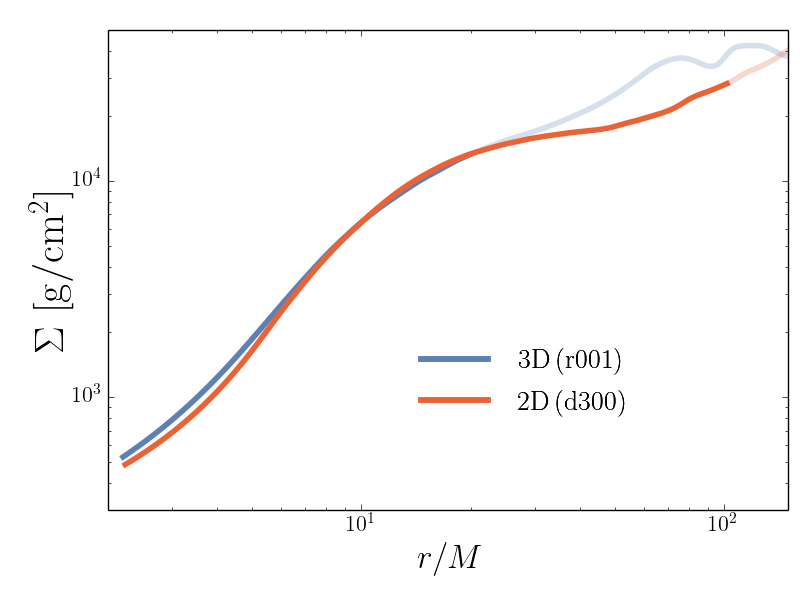}\vspace{-.95cm}
\includegraphics[width=1.\columnwidth]{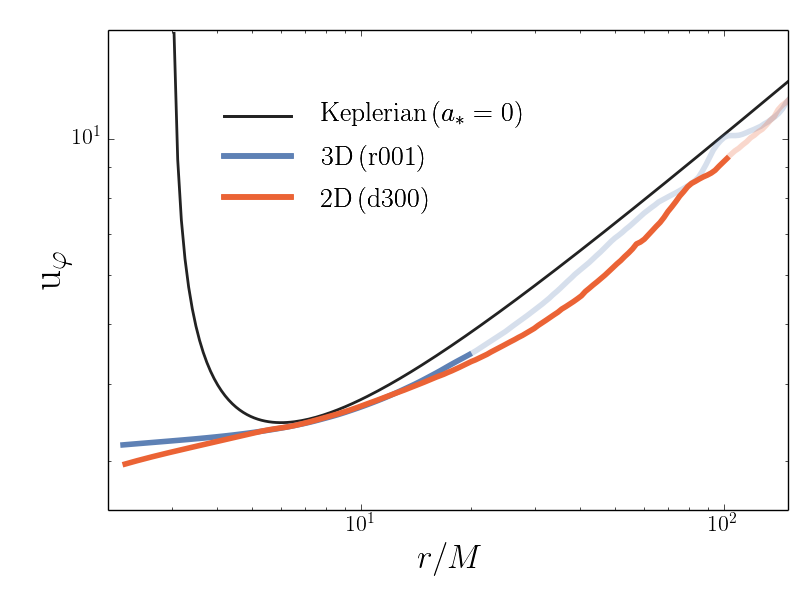}
\caption{Radial profiles of the net and outflowing accretion rate (top
  panel), surface density (middle), and specific angular momentum
  (bottom) for the three-dimensional run \texttt{r001} and the
  equivalent two-dimensional run \texttt{d300}. The light shaded line
  segments in the second and third panels correspond to regions where
  inflow/outflow equilibrum has not been reached.}
\label{f.rad2dvs3d}
\end{figure}

To quantify the comparison between models \texttt{r001} and
\texttt{d300}, we have calculated radial profiles of the accretion
rate, surface density, and density-weighted angular momentum for the
two runs. They are shown respectively in the top, middle, and bottom
panels in Fig.~\ref{f.rad2dvs3d}. The net accretion rates are
similar. The two-dimensional simulation has $\dot M=8.9\Medd$, roughly
$10\%$ lower than the three-dimensional model, but close enough to
allow a meaningful comparison of the two models. The extent of the
flat section of the net accretion rate profile is a benchmark for the
extent of inflow/outflow equilibrium. As discussed earlier, for
simulation \texttt{r001} the equilibrium region extends only to $r\sim
20$. Thanks to the long duration of the \texttt{d300} simulation, the
equilibrium region here extends much further out, to $r\sim
100$. Dashed lines in the same panel show the rate of mass lost in the
wind. It is comparable for the two runs, and suggests that the amount
of outflowing gas equals the net accretion rate around radius $r=60$.

The middle panel in Fig.~\ref{f.rad2dvs3d} compares the surface
density profiles. They are indistinguishable where the equilibrium
regions of the two simulations overlap (the sections of the curves
outside the equilibria are marked with shaded colors). Similarly, the
profiles of angular momentum also match well (bottom panel) except in
the plunging region inside the marginally stable orbit, where the
two-dimensional run results in a slightly lower angular momentum. This
difference probably arises from the fact that the dynamics of the flow
and the properties of the magnetic field in this region are not
governed by the combined effects of shear and the dynamo process
\citep{penna+alpha}.

In summary, the above comparisons indicate that the average properties
of the two- and three-dimensional simulations are remarkably
similar. The temporal properties, on the other hand, are significantly
different. In Fig.~\ref{f.lumvst_2d3d} we show radiative light curves
measured at angle $\theta=2^\circ$ and at radius $r=250$ (the light
curve of simulation \texttt{r001} corresponds to the one shown in the
top panel of Fig.~\ref{f.fluxvst_r020} for this particular
radius). The two-dimensional run resulted in noticeably larger
radiative luminosity in the outflow region and larger beaming towards
the axis (but the same total luminosity, see Table
\ref{t.luminosities}). As a result, the flux in the axisymmetrical
model is twice as large, and at the same time, it is also more
variable -- its fractional variability ($f=0.64$) significantly
exceeds that of the three-dimensional run ($f=0.23$). The difference
must come from the fact that the axisymmetrical simulation neglects
the many non-symmetrical modes which develop in three-dimensional
simulations. Variations in independent patches in azimuth would tend to
wash each other out when averaged, but no such averaging would be present
in an axisymmetric model.

\begin{figure}
\includegraphics[width=1.\columnwidth]{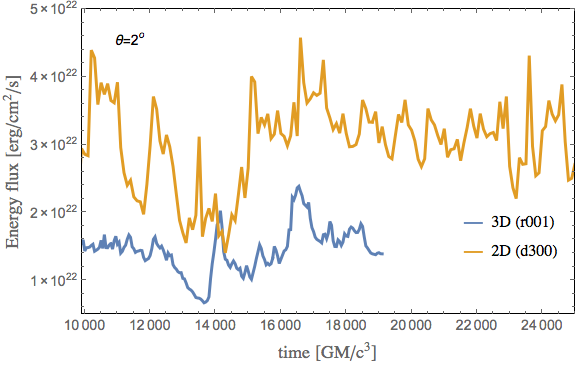}
\caption{Time variability of the radiative flux measured at angle
  $\theta=2^\circ$ from the axis at radius $r=250$ for the
  three-dimensional simulation \texttt{r001} and the two-dimensional
  model \texttt{d300}.}
\label{f.lumvst_2d3d}
\end{figure}

\section{Summary}
\label{s.discussion}

In this paper we presented a set of four three-dimensional simulations
of black hole super-critical accretion disks. Two of the simulations
accreted roughly at $10\medd$ (models \texttt{r001} and
\texttt{r020}), the third simulation (\texttt{r003}) had a
significantly larger accretion rate of $176\medd$, and the fourth
simulation (\texttt{r011}) was characterized by a rotating BH with
spin $a_*=0.7$ and accreted roughly at $17\medd$. The fifth simulation
we presented was performed assuming axisymmetry (hence this was a
two-dimensional simulation) and corresponded to roughly $10\medd$
accretion rate. All the simulations resulted in optically thick and
geometrically thick turbulent accretion disks.  In the course of this
study we reached a number of conclusions:

\begin{enumerate}
\item \textit{Photosphere:} Only for accretion rates $\dot M\lesssim
  10\medd$ does the photosphere extend down to the horizon, i.e., only
  for such low accretion rates will an observer at infinity viewing
  down the axis be able to see the innermost region of the accretion
  disk. For larger accretion rates, an on-axis observer will only
  observe a photosphere which is located relatively far from the black
  hole.  In fact, even for relatively low accretion rates $\dot
  M\lesssim 10\medd$, the optically thin region is limited to polar
  angles $\theta\lesssim 15^\circ$. For larger inclinations, an
  observer would only see the photosphere of the accretion disk
  located at large distance from the BH, often outside the
  computational box of the simulations. The limited box size and
  duration of the simulations do not allow us to study the radiation
  coming from such distant photospheres in any detail. We expect,
  however, that the spectrum of any such radiation will be relatively
  soft, because of the large distance from the BH, and that the
  observed isotropic-equivalent luminosity will not significantly
  exceed the Eddington value -- a super-Eddington flux in an optically
  thick wind causes acceleration of the wind and reduces the
  radiative flux towards the Eddington limit.

\item \textit{Stagnation radius:} For the simulations with the lowest
  accretion rates ($\dot M\lesssim 10\Medd$), the stagnation radius in
  the funnel, $r_0$, which separates an inner region where gas falls
  into the BH from an outer region where gas flows out, is located
  near the BH ($r_0\lesssim 10$). In the case of simulation
  \texttt{r011} (with a spinning BH), where the extraction of the
  rotational energy of the BH increases the energy flux through the
  funnel, the stagnation radius is very close to the horizon. For the
  run with the largest accretion rate (\texttt{r003}), the stagnation
  radius moves significantly out -- all gas in the funnel within
  $r_0\approx 50$ falls on the BH. This is a result of the increased
  optical depth of the gas, which traps photons and effectively
  suppresses the outflow of radiation up to this radius.

\item \textit{Total luminosity:} All the simulations with non-rotating
  BHs show the same total efficiency of roughly $3\%\dot M c^2$, approximately a factor of two
less than the efficiency of a standard thin accretion disk. We obtain
similar ratio of efficiencies for the simulation with a rotating BH ($a_*=0.7$),
for which the measured total efficiency was  $\sim 8\%\dot M c^2$. These
  efficiencies were calculated from the total luminosity in all
  forms of energy: radiative, kinetic, magnetic, thermal and binding
  (only rest mass energy was not included). The total luminosity is
  thus fundamental and represents the total energy
  deposited at ``infinity'' (the interstellar medium). There is no
  unique way of decomposing the above total luminosity into its
  constituent parts because of the limited size of the inflow/outflow
  equilibrium region in the simulations. When measured at the limit of
  inflow-outflow equilibrium ($r\approx 25$ in the disk interior),
  energy balance is dominated by the binding energy flux. This energy
  must be transformed into other forms of energy before reaching
  infinity (where the binding energy is zero).

\item \textit{Radiative luminosity:} Radiative luminosity can be
  measured reliably only inside the optically thin funnel region near
  the axis; here, radiation is decoupled from gas, and whatever
  radiation is flowing outward is guaranteed to reach the
  observer. However, only the simulations with the lowest accretion
  rates show an optically thin region inside the computational domain
  at all. Even in these cases, the optically thin radiative luminosity
  increases with radius (Table~\ref{t.luminosities}) because of
  radiation flowing into the funnel from the disk at larger radii. To
  obtain a useful estimate of the net radiative luminosity from the
  funnel, we would have to simulate accretion flows in much bigger
  computational box and for a much longer time. This is presently
  impractical. The radiative luminosities measured at radius $r=250$
  in the optically thin and outflowing regions are quite low. For
  accretion rates near $10\medd$ and a non-rotating BH, only $\sim
  20\%$ of the total liberated energy of $3\%\dot{M}c^2$ (mentioned
  above) comes out as optically thin radiation escaping through the
  funnel. The kinetic luminosity exceeds the radiative luminosity for
  the largest accretion rate considered, where the coupling between
  radiation and gas in the funnel is strongest, resulting in radiative
  acceleration of gas \citep{sadowski+radjets}. The luminosities are
  significantly larger for the simulation with a rotating BH because
  the BH spin provides an extra source of energy.

\item \textit{Beaming:} We have confirmed that radiation is beamed
  along the polar axis and the radiative fluxes here can be
  significantly super-Eddington. For the fiducial accretion rate of
  $10\Medd$, the simulations with non-rotating and spinning BHs show
  radiative fluxes of $20$ and $100 F_{\rm Edd}$, respectively, on the
  axis.  In contrast, the kinetic energy is not beamed on the axis but
  peaks (Fig.~\ref{f.fluxvsth}) either in a conical shell (for lower
  accretion rates) or in the wind region (for larger accretion rates).

\item \textit{Photon trapping:} We compared the total flux of
  radiation with the diffusive flux and showed that, for the accretion
  rates considered, advection of radiation (photon trapping) dominates
  over diffusive transport deep inside the disk. For $\dot M\approx
  10\Medd$, the advective flux is more than $30$ times stronger than
  the diffusive flux near the equatorial plane, and this ratio
  increases by at least an order of magnitude when the accretion rate
  grows to $\sim 176\Medd$, which is explained by the increased
  optical depth.  Closer to the disk surface or funnel boundary, the
  diffusive flux of radiation dominates, which ultimately contributes
  to the optically thin radiation escaping through the funnel. This
  flux also helps to drive gas away from the disk surface into the
  funnel.

\item \textit{Radiation transport} An analysis of our simulations
  shows that radiation transport is dominated by advective and
  diffusive transport. We do not see a significant component of
  turbulent radiative transport.

\item \textit{Advection factor:} For the fiducial model ($\dot
  M=10\medd$) we estimated the fraction of photons generated in the
  disk that manage to escape vertically and enter the optically thin
  funnel region. At radius $r=10$, only $ \sim 20\%$ of photons leave
  the disk while $\sim80\%$ end up in the BH. Corresponding numbers at
  radius $r=25$ are approximately $45\%$ and $55\%$,
  respectively. This suggests that the effective trapping radius,
  where half the photons escape from the disk and half fall into the
  BH, is located at $r\gtrsim 30$ for $10\Medd$.

\item \textit{Variability:} We attempted to extract from the
  simulations (frequency integrated) lightcurves as seen by observers
  at infinity. Because of the limited size of the equilibrium region,
  and the fact that the photosphere is located close to the axis even
  for the lowest accretion rate considered, we were able to obtain
  robust light curves only for radiation escaping along the axis, and
  that too only for the fiducial model. Extracting variability
  information from simulations of optically thick and geometrically
  very thick disks is, and will remain to be, challenging. One may
  expect that thin disks, with photospheres much closer to the
  equatorial plane, will be more amenable to study.

\item \textit{Impact of the BH mass:} We performed one simulation
  (\texttt{r020}) with BH mass $1000\Msun$, which had the same
  Eddington-scaled mass accretion rate as the fiducial model
  (\texttt{r001}). The properties of the two simulations, after
  scaling down the latter by appropriate factors, were quite similar;
  in fact, it was difficult to distinguish the two (compare
  Fig.~\ref{f.5panels}). This is because electron scattering opacity
  dominates in both simulations, and under these conditions the
  accretion equations scale very simply with BH mass.  For much larger
  BH masses (corresponding to the SMBH regime), the absorption opacity
  will no longer be negligible and this will break exact scaling.

\item \textit{Axisymmetric simulations:} We compared the properties of
  the fiducial three-dimensional simulation (\texttt{r001}) with a
  two-dimensional axisymmetric simulation (\texttt{d300}) that made
  use of an artificial magnetic dynamo and accreted at roughly the
  same rate. We showed that the time-averaged properties of the two
  simulations were remarkably similar. Because of the significantly
  lower computational cost, we were able to run the axisymmetric
  simulation for a much longer time and obtained a significantly
  larger equilibrium region (extending up to $r\sim100$ instead of
  only $r\sim25$ in the case of the 3D model). Thus, two-dimensional
  axisymmetric simulations (with magnetic dynamo) are a cheap and
  promising method for running models for long times and thereby
  extending the range of radii over which one obtains useful
  information. However, this must be done with caution.  While
  time-averged quantities appear to be reliable, we note that the
  temporal properties of the 2D and 3D runs were different. In
  particular, because of the lack of non-axisymmetric modes in the
  axisymmetric simulation, it showed much larger variability in the
  radiative flux flowing out along the axis.

\end{enumerate}

\subsection{Comparison with other studies}
\label{s.comparison}

Our study is the first to explore the parameter space relevant to
radiation-dominated BH accretion disks using three-dimensional,
general relativistic, radiation-MHD simulations. The properties of the
simulated super-critical disks described in this work are
qualitatively in agreement with previously published global
simulations.  In particular, significant photon trapping was
identified already by \cite{ohsuga07}, and confirmed more recently by
\cite{sadowski+koral2} and \cite{mckinney+harmrad}. The total
luminosities given in the present work are also close to those
obtained in the latter studies. The same is true for radiation beaming
and the radiative luminosity of the funnel.

There are a number of differences between our results and those
described in \cite{jiang+14b}. The model presented in the latter work,
which had an accretion rate of $\sim 15\Medd$, can be directly
compared to our fiducial model \texttt{r001}. \cite{jiang+14b} found a
very different spatial distribution of gas compared to our run; their
disk is strongly concentrated at the equatorial plane whereas our disk
is not. Perhaps because of this, they obtained a powerful radiative
flux from the innermost ($r<10$) region; in fact, nearly all of their
radiative luminosity, which is of the order of $5\%\dot M c^2$, comes
from such small radii.  Our simulations show a similar total
efficiency ($3\%\dot M c^2$), but only a small fraction of the energy
escapes as radiation at small radii.

\cite{jiang+14b} used a Newtonian code with cylindrical coordinates
(and a ``cylindrical BH'' with radius $r=4$). Their simulation was
intialized with a different and more strongly magnetized torus
compared to ours. They implemented a sophisticated radiation transfer
solver which evolved in real time a number of specific intensities. In
comparison, our simulations were done with the M1 closure scheme,
which means that we evolved only four independent radiation quantities
in each cell.  On the other hand, \cite{jiang+14b}, for the sake of
performance, had to make some approximations when treating the
interactions between gas and radiation\footnote{These approximations
  were avoided by \cite{ohsugatakahashi-15}, who implemented
  sub-stepping in their implicit solver, but otherwise used the same
  ideas as in \cite{jiang+14b}.}. Which of these factors is
responsible for the large discrepancy between their study and ours is
presently unclear.

\section{Acknowledgements}

The authors thank Jean-Pierre Lasota for his comments.  AS
acknowledges support for this work by NASA through Einstein
Postdoctotral Fellowship number PF4-150126 awarded by the Chandra
X-ray Center, which is operated by the Smithsonian Astrophysical
Observatory for NASA under contract NAS8-03060. AS thanks
Harvard-Smithsonian Center for Astrophysics for hospitality.  RN
was supported in part by NSF grant AST1312651 and NASA grant TCAN
NNX14AB47G.  The authors acknowledge computational support from NSF
via XSEDE resources (grant TG-AST080026N), and from NASA via the
High-End Computing (HEC) Program through the NASA Advanced
Supercomputing (NAS) Division at Ames Research Center.
 
\bibliographystyle{mn2e}

\end{document}